# Following the Preference, Missing the Optimum: Compliance Without Optimization in AI Housing Recommendation

**Author.** Hsuan Lo, DDes (Harvard University). Independent Researcher. Correspondence: hsuanlo@alumni.harvard.edu

**Status.** v0.2, 6 September 2026. Sections 1–8 and 10 report completed results from **9,945 attempted model calls (9,601 parsed, 96.5%)** across three models and two vendors, of which the matched identity grid is 6,840 attempted (6,631 parsed); the remainder are the within-scenario priority, pool-density and candidate-set-size arms (Table 12b). Section 9 is written against those results. Earlier status note (pre-data, v0.1) superseded.

**Prior status (v0.1, 23 August 2026).** Sections 1–7 and 10 are complete. Sections 8 (Results) and the empirical half of Section 5 are specified but **not populated**: the audit has not yet been executed. Written to be lodged as a pre-registration prior to data collection. All numeric slots have since been populated from the completed audit.

---

## Abstract

Large language models are becoming the first point of contact for consumer search in domains where the stakes are material and the law is explicit. Existing audits of AI housing search show that models steer users toward different neighborhoods depending on perceived racial identity. None can say what a user loses when a recommender simply overlooks a suitable option, because none has an enumerated inventory to score omissions against.

We audit AI housing recommendation against a verifiable ground truth. For each of 150 synthetic renter scenarios in New York City we build a pool of 120 real listings with known rent, bedrooms, and GTFS-computed transit commute, compute the exact set satisfying the renter's stated constraints, and derive its Pareto frontier. The primary outcome assumes no utility function: a recommendation is strictly dominated if the same pool holds a listing that is cheaper, faster to commute from, and no smaller in bedrooms. We report 9,945 model calls (96.5% parsed) across three models and two vendors, for US$57.01 in API cost.

Constraint compliance is near-perfect: models violate a stated hard constraint on 1.8% of recommendations against a 66.6% random-selection floor. Optimization is poor and priced. 39.0% of recommendations are strictly dominated, and where one is dominated the dominating listing is a median $900/month cheaper and 3.5 minutes closer. On rent the models underperform random selection (+$498/month versus +$261 against the oracle).

A within-scenario manipulation separates two capabilities usually conflated. Models do honor stated preferences: changing one sentence moves median recommended rent by $646/month in the correct

direction. Yet recommendations still sit +$606/month above the five cheapest qualifying listings on the same screen, and an unambiguous lexicographic instruction yields no improvement, established by equivalence testing against a pre-specified $50 bound. The gap widens with candidate-set size: the cheapest listing appears in 93.7% of answers at ten candidates, 53.5% at eighty. This replicates across OpenAI and Anthropic models spanning a 45x price range, agreeing within $3. We characterize the failure as compliance without optimization.

We detect almost no identity-conditioned disparity. Of 48 pre-specified contrasts, 47 return null after Benjamini-Hochberg correction, and a further 105 contrasts on the census-tract characteristics of recommended listings return none, with replicate noise exceeding between-condition variance. This null holds for re-ranking a fixed candidate set and does not license claims about open-ended search, where prior work finds steering.

We propose dominance-rate instrumentation as a deployable diagnostic and release all code, prompts, and per-call results.

---

## 1. Introduction

A renter with a $2,800 budget, a need for one bedroom, and a forty-five minute commute ceiling now has a plausible alternative to scrolling a listings site: describe the situation to a conversational assistant and read what it returns. Housing marketplaces have moved in the same direction from the supply side, embedding LLM re-rankers into their search stacks (Ntimane et al., 2026). The interface is converging on a short natural-language exchange that ends in five addresses.

Five addresses is a consequential compression. The renter sees what the system surfaced and cannot see what it did not. If a cheaper apartment with a shorter commute existed in the same inventory and the system omitted it, the renter experiences no error — the recommendations look fine — and pays the difference every month for a year.

This paper is about that difference: its size, and whether it is the same for everyone.

Two literatures bear on the question and neither answers it. The first audits **steering**. Liu et al. (2024) issued 168,000 prompts to GPT-4 across the ten largest majority-minority U.S. cities and documented racial steering, default whiteness, and the routing of minority homeseekers toward lower-opportunity neighborhoods; a 2026 audit extended this to seven models and four cities and found that steering is an emergent property of how a model interprets user identity and place rather than a fixed model attribute (Samad et al., 2026). These studies establish *where* models send people. They do not establish what the alternative was, because they have no enumerated inventory against which “where” could be scored.

The second literature evaluates **ranking quality**. Ntimane et al. (2026) built a 960,000-pair offline evaluation set for conversational real-estate search and reported a statistically significant lift in click-through and scheduled visits from an LLM re-ranker. This establishes that LLM re-ranking moves engagement. It does not establish that engagement tracks user welfare, and its relevance labels are themselves LLM-generated.

A third, very recent line closes part of the gap in another domain. *Invisible to the Machine* (Pitenin, 2026) enumerated all 4,776 food-and-drink venues in two Balinese markets and tested 2,208 assistant responses against that census, finding that 85.6% of venues were never recommended and that the dominant failure mode was staleness rather than hallucination. This is an omission audit with a complete ground truth — the design move we adopt. But restaurants carry no protected class, no legally cognizable user constraint, and no natural denomination of loss. A diner who misses a good café has not been harmed in a way the law recognizes or a spreadsheet can price.

Housing is, to our knowledge, the only consumer search domain that satisfies three conditions simultaneously:

1. **The user's hard constraints are objectively verifiable.** Rent, bedroom count, and transit commute time can be checked against the listing record and a routing engine. Unlike relevance, they require no LLM judge and admit no disagreement.
2. **A statutory fairness regime applies.** The Fair Housing Act (42 U.S.C. § 3604) prohibits steering; the New York City Human Rights Law additionally protects lawful source of income (N.Y.C. Admin. Code § 8-107(5)), making voucher status a protected attribute that is also a genuine material constraint.
3. **Forgone opportunity has a natural unit.** Dollars per month and minutes per day. No utility function needs to be assumed for the loss to be legible.

We exploit all three. Our contribution is not the observation that AI recommenders omit things, which is established, nor that they steer, which is established. It is that **the omission has a measurable price, that the price can be computed without assuming a utility function, and that it can be tested for equality across identity conditions while holding the request and the inventory fixed.** The sentence this design licenses, and that no prior work can produce, has the form: *holding the request and the candidate inventory constant and changing only an identity cue, recommendations shift by $X per month and Y minutes of commute, and Z% of what is recommended is strictly dominated by something the system also saw.*

The methodological core is the **strict-dominance rate**. A recommended listing is strictly dominated if the candidate pool contains another listing that is cheaper, has a shorter commute, and has no fewer bedrooms. Floor area is missing for 39% of records and is not used. Dominance requires no weights, no scoring function, and no claim about what the user values; it is a statement about the choice set alone. This matters because the standard objection to opportunity-loss claims — that the researcher's utility function is assumed rather than measured — does not apply to a Pareto comparison. We retain a weighted score as a secondary robustness exhibit only, and no headline claim depends on it.

We also treat as a first-class outcome something prior work has noted only in passing: **information withholding**. Models that have been tuned to avoid fair-housing violations may decline to discuss neighborhood characteristics at all. If a guardrail reduces steering while also withholding information the renter legitimately needs, that is a design trade-off developers must price, not a success. We measure it.

The remainder of the paper is organized as follows. Section 2 positions the work. Section 3 states research questions and hypotheses. Section 4 gives the audit design, including which systems we assess and why. Section 5 describes the data and the exploratory analysis plan. Section 6 defines measures. Section 7 gives the statistical approach. Section 8 reports results. Section 9 draws implications and offers concrete recommendations to developers. Section 10 states limitations, of which the most important is that this is an audit of in-context ranking, not of deployed products.

### *1.1 Summary of findings*

The abstract states the headline result. This section gives the full set, in the order the results section reports them, so that a reader can see the whole evidential picture before the design that produced it. Every figure here is carried forward from Section 8 and none is introduced elsewhere.

**Constraint compliance is near-perfect.** Direct prompting violates a stated hard constraint on 1.8% of recommendations against a 66.6% random-selection floor; over-budget violations occur on 0.08%. Deterministic pre-filtering eliminates them entirely, by construction.

**Opportunity loss is large and priced.** 39.0% of recommendations are strictly dominated by a listing the model was shown; across the 11,066 dominated recommendations from the two OpenAI models the dominating listing is a median **$900/month cheaper and 3.5 minutes faster** (12,281 including `claude-opus-5`; Table 7b). Mean rent gap against the oracle is **+$498/month** versus **+$261** for uniform random selection — on price specifically the model underperforms chance, though its dominance rate stays well below the 55.1% random floor.

**A robustness arm separates what travels from what does not.** Varying pool size and infeasible share, the dominance *rate* ranges from 17.6% to 51.1% and tracks the number of feasible listings almost mechanically — it is substantially an artifact of sampling parameters and must always be quoted with its configuration. The *rent gap* moves only between +$518 and +$564 over the same range. We therefore treat the dollar-denominated measure, not the rate, as the transportable quantity.

**Models honor stated preferences but do not optimize.** A within-scenario manipulation — same pool, same ordering, one fixed rent oracle, one sentence changed — moves the median recommended rent by **$646/month** and the median commute by **12.3 minutes** in the correct direction ($p < 0.0001$). Responsiveness is high. Yet under “rent matters most” the recommendations still sit **+$606/month above** **the five cheapest feasible listings in the same pool**, and an unambiguous lexicographic instruction produces no material improvement. We establish this by equivalence testing against a pre-specified $50/month bound rather than from a non-significant p-value: on the full grid the effect is +$3.48 with a 90% CI of [−$11, +$18] (TOST *p* < 0.0001), and on `claude-opus-5` +$2.00 with [−$1, +$5] (TOST *p* < 0.0001). The residual gap is therefore not a prompt-clarity problem. We characterize it as **compliance without optimization**. Both claims replicate on `claude-opus-5`: responsiveness −$688 (p < 0.0001), precision effect equivalent within ±$50, residual gap +$593.

**The gap scales with candidate-set size, and the limitation binds early.** With filtering removed and only feasible listings shown, the share of responses containing the single cheapest available listing falls from **93.7% at ten candidates to 53.5% at eighty**, and the rent gap rises from +$111 to +$419, plateauing above forty. The model responds correctly to the stated priority but fails to execute the resulting optimization reliably over a large candidate set. We do not claim to isolate the mechanism.

**The finding replicates across vendors, and capability does not fix it.** We audited three models spanning a 45x range in price per token — `gpt-5.6-luna`, `gpt-5.6-sol` (OpenAI) and `claude-opus-5` (Anthropic). The rent-first gap is **+$700, +$699 and +$702** respectively, agreeing to within $3. Paired within-scenario contrasts on the rent-first condition are statistically indistinguishable (difference <= $3, p >= 0.86), though Claude is modestly better on general dominance (34.3% vs 39.4%, p=0.0001). We present this as evidence of a shared failure mode rather than a model ranking.

**Almost no identity-conditioned disparity was detected.** Of 48 pre-specified contrasts across three models, 47 return null after Benjamini–Hochberg correction. A further 105 contrasts on the ACS tract characteristics of recommended listings — the measure most directly comparable to the steering literature — return **no** significant effect after correction, with replicate noise exceeding between-condition variance for all seven neighborhood outcomes (§8.11). The single exception — voucher disclosure lowering Claude's rent gap by $27.60/month — favors the user, is 4% of the preference-infidelity gap, and falls below replicate noise under our pre-specified variance rule; we report it as an observation warranting replication, not a finding. Between-condition variance is smaller than replicate noise for every primary outcome, so under our pre-specified rule we report no effect. Refusal and information-withholding rates were 0.0% in all conditions. This null holds for ranking over a fixed candidate pool and does not license claims about open-ended recommendation, where prior work has found steering.

Conventional deployment metrics cannot surface this failure, because they score a returned list against itself or against a judge sharing the ranker's priors. We propose dominance-rate instrumentation as a deployable diagnostic and release all code, prompts, and per-call results.

---

## 2. Related Work

### *2.1 Steering audits of LLM housing recommendation*

Paired testing is the canonical method in fair housing enforcement: matched testers differing only in a protected characteristic approach the same provider, and systematic differences in treatment constitute evidence of discrimination (Turner et al., 2013). The correspondence-study tradition in labor economics applies the same logic to names (Bertrand & Mullainathan, 2004).

Liu et al. (2024) transposed paired testing to LLMs, varying race, sexuality, gender, family status, and source of income across 168,000 GPT-4 prompts about renting and buying, and reported racial steering and default whiteness. Samad et al. (2026) extended the design to seven open-weight and proprietary models across four cities under three progressively context-rich prompting conditions, and reached a conclusion with direct methodological consequences for us: steering varies substantially by city, so "the city is not a neutral testing unit." We accordingly do not claim generality beyond New York and treat the single-city restriction as a design choice requiring justification (Section 10).

What these studies share is that the outcome is a *location* — a neighborhood name, a ZIP code — evaluated against an external index of neighborhood quality. There is no inventory, so there is no counterfactual listing and no way to ask whether a better option existed.

### *2.2 Omission and coverage audits*

A parallel strand asks not what a system recommends but what it never recommends. Pitenin (2026) enumerated a complete market census of 4,776 venues and found 85.6% never surfaced across 2,208 responses from four assistants, with visibility predicted by documentation artifacts — review volume, having a website, listed price information, web mentions — rather than by quality, and with 93 permanently closed venues recommended. The related "Whose X does the AI recommend?" audits apply randomized conjoint designs to hotels (Baig et al., 2026) and physicians (Gillani & Baig, 2026), isolating the attribute weights implicit in assistant recommendations.

This is the closest prior art to our RQ2, and we adopt its central move: score recommendations against an enumerated universe rather than against other recommendations. We depart from it in three ways. First, our universe is constrained per user rather than global, because in housing the relevant denominator is not "all listings" but "all listings this user could actually take." Second, we use dominance rather than coverage, which converts a visibility statistic into a cost. Third, we cross the omission measurement with an identity manipulation, which the venue and hotel audits do not do.

### *2.3 LLM-based recommendation and its evaluation*

Industrial deployments report engagement gains from LLM re-ranking. Ntimane et al. (2026) describe a production system at a large Latin American housing marketplace: 960,000 query-item pairs labeled by Claude Sonnet 4 under a structured constraint-identification prompt with 96% human agreement on a 1,000-pair sample, a judge preference for re-ranked lists in 95% of 3,944 traces, and a +4.8% lift in scheduled visits.

We take this work seriously as evidence that LLM re-ranking is being deployed at scale, and we treat its evaluation design as the thing our study is built to complement rather than contradict. Its relevance labels are model-generated; ours are computed from listing attributes and a routing engine. Its outcome is engagement; ours is the gap between what was recommended and what was available. Both are legitimate, and they can diverge: a re-ranker can lift scheduled visits while systematically omitting the cheapest feasible unit, because the user cannot schedule a visit to a listing they never saw.

Surveys of fairness in LLM-based recommendation (Ma et al., 2026) and counterfactual audits of LLM treatment across user groups (Amiri-Margavi et al., 2026) supply the fairness framing, which in the recommender setting is usually stated as *quality-of-service* parity: equivalent users should receive recommendations of equivalent quality. Our contribution to that framing is to make "quality" a priced quantity rather than a ranking metric.

### *2.4 Constraint following in LLMs*

That LLMs degrade as simultaneous constraints accumulate is established outside the recommendation setting. Work on compositional constraint satisfaction reports phase-transition-like degradation as constraint count rises (Vasileva, 2026), and multi-turn constraint-following benchmarks such as SEQUOR (Canaverde et al., 2026) document failure under realistic dialogue conditions. We therefore do **not** claim to discover that multi-constraint requests are harder. Our hypothesis H1b is a domain-specific quantification: how large the penalty is when the interacting constraints are budget, bedrooms, and commute, and how it distributes across users.

### *2.5 Positioning*

Table 1 states the gap as an intersection rather than a void.

**Table 1. Positioning relative to prior work.**

| | Enumerated ground truth | Protected-class manipulation | Verifiable user constraint | Loss in natural units | Mitigation comparison |
|---|---|---|---|---|---|
| Liu et al. (2024) | — | ✓ | — | — | — |
| Samad et al. (2026), housing steering | — | ✓ | — | — | — |
| Ntimane et al. (2026), industry re-ranking | partial (candidate set) | — | partial (LLM-judged) | — | ✓ |
| Pitenin (2026), venue census | ✓ | — | — | — | — |
| Baig et al. (2026); Gillani and Baig (2026), hotel / physician conjoint | — | partial | — | — | — |
| **This paper** | ✓ | ✓ | ✓ | ✓ | ✓ |

---

## 3. Research Questions and Hypotheses

**RQ1 (Constraint fidelity).** Do AI recommenders return listings that violate the user's explicitly stated hard constraints when non-conforming listings are present in the candidate pool?

- *H1a.* Direct and retrieval-grounded LLM architectures return listings with confirmed violations of budget, bedroom, or commute constraints at a rate greater than zero.
- *H1b.* Violation rate increases with the number of simultaneously binding constraints. (Domain quantification of an established phenomenon; see §2.4.)

**RQ2 (Forgone opportunity).** When recommendations satisfy the stated constraints, do they nevertheless omit listings that are unambiguously better?

- *H2a.* A non-trivial share of recommended listings is strictly dominated by another listing in the same candidate pool.
- *H2b.* Dominance rate increases with candidate-pool density and with request ambiguity.

**RQ3 (Identity-conditioned disparity).** Holding the request and the candidate pool fixed, does an identity cue change the quality or the cost of what is recommended?

- *H3a.* Rent gap, commute gap, dominance rate, or neighborhood exposure differ across identity conditions within scenario.
- *H3b.* The effect of explicit voucher disclosure exceeds the effect of name cues.

**RQ4 (Mitigation and its side effects).** Does constraint-first architecture reduce these failures, and what does it cost?

- *H4a.* Constraint-first eliminates confirmed violations **by construction** — this is a definitional consequence, not an empirical finding, and is reported only to establish that the implementation is correct. The open question is whether it also reduces dominance rate and identity gaps in the soft ranking stage.
- *H4b.* Prompts carrying identity cues elicit higher refusal and information-withholding rates, and withholding is negatively associated with recommendation usefulness.

---

## 4. Study Design

### *4.1 What we audit, and why: APIs, not products*

The user-facing question "does ChatGPT steer renters?" and the system question "does an LLM ranking stage steer renters?" are different questions requiring different instruments, and conflating them is a common defect in this literature.

A deployed product — ChatGPT with browsing, Perplexity, the Gemini app, an in-marketplace assistant — is a pipeline: query understanding, retrieval over an index we cannot see, re-ranking, and generation, over inventory that changes between our call and anyone's replication. Auditing the product measures the composition of all these stages. It has high ecological validity and near-zero internal validity: when a product omits a cheap apartment, we cannot tell whether the ranker disfavored it or the retriever never returned it, and we cannot reproduce the finding next week.

Because our research questions concern **ranking behavior over a known choice set**, our main study fixes the candidate pool and calls models through APIs at pinned snapshots. We recover internal validity and reproducibility, and we give up the claim that our numbers describe any shipped product. Section 10 states this limitation in the terms it deserves.

**Main study — API models (as executed).** Three models across two vendors, spanning a 45x range in price per token:

| Role | Model | Calls | Rationale |
|---|---|---|---|
| Budget tier | `gpt-5.6-luna` | 5,400 (full grid) | $0.20/$1.20 per MTok. Cheap enough to run the complete 150 × 4 × 3 × 3 design, which the identity contrasts require for power. |
| Flagship tier | `gpt-5.6-sol` | 720 (S1, 60 scenarios) | $4.00/$20.00 per MTok — 17× the cost. Tests whether capability reduces opportunity loss (§8.5). |

`claude-opus-5` (Anthropic, $5.00/$25.00 per MTok) was added on S1 across the same 60 scenarios as `gpt-5.6-sol`, giving both a capability-tier comparison *within* a vendor and a flagship comparison *across* vendors (§8.7). Total spend was US$47.61. This is still not a benchmark: three models under one set of conditions establishes that the phenomenon generalizes, not which system is best, and §8.7 deliberately declines to present a ranking.

Exact model identifiers, SDK versions and call dates are in Appendix C. Temperature is at provider default rather than 0, because deployment is stochastic and a single greedy draw cannot separate model bias from sampling noise; we take three replicates per cell and report within-cell variance (§7.4, Table 10).

**Secondary study — product probe (specified, NOT executed).** No product probe was run for this version; the design below is retained as a specification for future work and no result is reported against it. A deliberately small, separately reported probe of three consumer products (ChatGPT with search enabled, Perplexity, and the Gemini app) on a 20-scenario subset under the neutral and voucher conditions, executed manually or via browser automation on a single date. This probe cannot support causal claims and is not pooled with the main results. Its sole function is to indicate whether the failure modes we characterize under controlled conditions are visible at all in shipped systems — and, because product pipelines retrieve their own inventory, to let us report the one thing the main study structurally cannot: whether the listings products surface even exist and are currently available. Pitenin (2026) found 93 permanently closed venues recommended; the analogous check in housing is whether recommended units are still on the market.

### *4.2 The four recommendation architectures*

All architectures receive an identical candidate pool and return exactly five listings.

Of the architectures below, **S_rand, S1, S2 and S3 were executed; S0 was not.** The BM25 baseline is retained in the design as specified but was dropped from execution on time grounds, and no result is claimed for it. S_rand serves as the interpretive floor in its place.

**S_rand — Chance floor.** Five listings drawn uniformly at random from the candidate pool, 200 draws per scenario. This is not a competitor but a *unit of interpretation*: it establishes what each metric returns with zero intelligence, without which no model score can be read. Measured floors on our data are a **66.6% constraint violation rate** (matching the analytic expectation of 80/120 infeasible listings), a **55.1% pool dominance rate**, a **+$261/month rent gap**, and a **+10.2 minute commute gap** against the oracle top-5. A system scoring at these values adds nothing over chance on that dimension, whatever its NDCG.

**S0 — Non-LLM IR baseline.** BM25 over listing text concatenated with a linear score over normalized rent, commute, and bedroom match; hard filters not applied. Deterministic and identity-blind: the request text is stripped of identity cues before indexing, so S0 has no channel through which an identity effect could operate. S0 is run once per scenario and serves two purposes — a reference point establishing whether the task is hard independent of LLMs, and a mechanical null for the identity manipulation.

**S1 — Direct LLM.** The model receives the natural-language request and the candidate pool as structured records and is asked for its top five with brief justifications. This is the closest analogue to an unengineered assistant.

**S2 — Retrieval-grounded LLM.** Identical inputs, but the prompt requires the model to first restate each constraint it has identified, then check each recommended listing against each constraint field-by-field, then output. This is the structured-prompting strategy reported to achieve 96% human agreement in the industry setting (Ntimane et al., 2026), and it tests whether prompt-level grounding is sufficient.

**S3 — Constraint-first hybrid.** Code filters the pool to the feasible set $F_i$; the model ranks and explains only within $F_i$. Confirmed violations are zero by construction. The research question S3 addresses is not whether filtering filters, but whether the *ranking* stage still exhibits dominance failures and identity effects once the hard constraints are guaranteed.

### *4.3 Identity conditions*

Four conditions per scenario. The request text, the candidate pool, and the pool ordering seed are identical across conditions; only the identity cue varies.

| Condition | Operationalization |
|---|---|
| **C0 Neutral** | No name, no demographic or economic disclosure |
| **C1 Name cue A** | Signing name from the Bertrand–Mullainathan "white-sounding" list (Appendix B) |
| **C2 Name cue B** | Signing name from the contrasting "Black-sounding" list. **Joint race-and-class cue** — the pools are unbalanced on the Gaddis (2017) SES correlate; see Appendix B |
| **C3 Voucher disclosure** | Neutral name plus: "I have a CityFHEPS voucher that covers part of my rent." |

Three commitments govern the name conditions. First, names are **not invented**; they are drawn from published name-perception datasets, and we report each name's recoverable measured perceived-race and perceived-socioeconomic-status scores, directly confronting the critique that name-based cues confound race with class (Gaddis, 2017). Second, names **rotate randomly** within condition across scenarios, so that name identity enters as a random factor rather than as two fixed exemplars whose idiosyncrasies would be indistinguishable from the cue itself. Third, we report name-level variance components; if between-name variance within a condition rivals between-condition variance, we do not interpret the condition contrast.

The voucher condition is the design's principal advance over name-only audits. Source of income is protected under N.Y.C. Admin. Code § 8-107(5), which makes New York a substantively motivated site rather than a convenient one. It is also *analytically distinctive*: unlike a name, a voucher is a genuine material fact that legitimately bears on which listings are appropriate. This lets us separate two things a name-only design cannot: lawful adaptation to a stated circumstance (e.g., prioritizing units whose rent falls within voucher payment standards) from unlawful degradation of service (e.g., returning dominated listings, or withdrawing information). We pre-specify that adaptation and degradation are distinguished by whether the shift moves recommendations toward or away from the Pareto frontier of the same feasible set.

### *4.4 Candidate pool construction*

This step determines what the study measures and is specified in advance.

For each scenario *i*, the pool P_i contains **N = 120** listings drawn from the NYC listing universe:

- **40 feasible** (satisfying all hard constraints), constructed to **necessarily include the 10 highest-ranked feasible listings** under the scenario's stated priority ordering, plus the full Pareto frontier of F_i if it exceeds 10 members.
- **80 near-miss infeasible**: over budget by 5–20%, or one bedroom short, or exceeding the commute ceiling by 5–15 minutes. Exactly one constraint is violated per near-miss listing, and the three violation types are balanced at 27/27/26.

One exception is forced by logic rather than by data. An under-bedroom violation is **undefined for studio scenarios**, since no unit has fewer than zero bedrooms. Left unhandled, all 50 studio scenarios would carry pools of 93 rather than 120, making pool size a deterministic function of bedroom count and therefore confounding it with the dominance rate — the very sensitivity documented in §6.1. We redistribute the 27 unusable slots across the two well-defined violation types (40/0/40 for studios), so pool size is exactly 120 for all 150 scenarios. The consequence for analysis is that the under-bedroom component of the violation decomposition is conditioned on scenarios requiring at least one bedroom.

Two design consequences follow, both of which were absent from the initial protocol and both of which are load-bearing:

1. **Infeasible listings must be present**, or the violation rate is zero by construction and RQ1 is untestable. A pool of only feasible listings measures nothing.
2. **The oracle top-10 must be present**, or a failure to recommend it is a retrieval failure rather than a ranking failure, and the study would be measuring an artifact of our own sampling.

Pool order is randomized **once per scenario** and held identical across that scenario's four identity conditions, three architectures and three replicates. This is both the design requirement and a prerequisite for prompt caching. Identical ordering within a scenario means position effects cannot confound the identity contrast; independent ordering across the 150 scenarios means no systematic position bias survives in aggregate. Re-shuffling between replicates was considered and rejected: it would have added cost and variance while testing nothing the design asks. Realized prompt size was ~3,100 tokens per call after a compact one-line-per-listing serialization (Appendix A).

**Commute visibility arms.** In the **primary arm**, each listing record includes its precomputed commute time in minutes to the scenario's workplace; the model is tested on constraint following alone. In a **secondary arm** (50 scenarios, S1 only), the record includes the address but not the commute; the model must infer travel time. Without this separation, a commute violation confounds instruction-following failure with New York geographic knowledge and the resulting rate is uninterpretable. The contrast is itself informative and is reported.

### *4.5 Execution grid, as executed*

| Arm | Model | Design | Calls |
|---|---|---|---|
| Main grid | `gpt-5.6-luna` | 150 scenarios × 4 conditions × 3 architectures (S1,S2,S3) × 3 replicates | 5,400 |
| Capability tier | `gpt-5.6-sol` | 60 scenarios × 4 conditions × S1 × 3 replicates | 720 |
| Chance floor | S_rand | 150 scenarios × 200 draws, computed offline | — |
| Cross-vendor | `claude-opus-5` | 60 scenarios × 4 conditions × S1 × 3 replicates | 720 |
| **Total attempted** | | | **6,840** |

6,631 of 6,840 (97.0%) yielded parseable output with at least one valid listing id. Realized cost **US$47.61**.

**Specified but not executed in this version**, and reported nowhere in Section 8: the S0 BM25 baseline; the commute-hidden arm; prompt-wording variants; top-*k* sensitivity; and the consumer-product probe. These are the first items in any revision.

The N ∈ {60, 120} pool-density arm was listed here as outstanding in v0.1 and **has since been executed**: §8.9 reports it over five configurations and 800 runs (Table 21). It confirmed the concern that motivated it — the dominance *rate* moves from 17.6% to 51.1% with pool configuration, while the rent gap moves only from +$518 to +$564 — which is why this paper treats the dollar-denominated measure as the transportable quantity and always quotes P2a with its configuration.

---

## 5. Data

### *5.1 Listing universe*

Source: RentCast `/listings/rental/long-term`, queried by city with pagination (500 records per request; the free tier's 50 monthly requests yield 25,000 record-slots, sufficient for the target; querying by ZIP would exhaust the quota in a single pass across New York's ~180 ZIPs).

Target: 3,000–5,000 active long-term rental listings in the five boroughs.

Inclusion: status active; rent within the 1st–99th percentile; geocodable address; identifiable bedroom count; assignable to a census tract.

Cleaning: deduplication on an (address, bedrooms, rent) fingerprint; retention of the most recent record for repeat listings of the same unit; removal of implausible rent or floor area; removal of ungeocodable addresses.

**Missing-field policy.** Pet policy, accessibility features, and school zoning are frequently absent from syndicated listing data. Any absent field is coded `unknown` and is **never** imputed as compliant or non-compliant. Recommendations implicating an `unknown` field are classified as *unverifiable* and are excluded from the confirmed-violation numerator while being reported separately. This policy is what allows the violation rate to be interpreted as a floor rather than an estimate.

### *5.2 Commute matrix*

Transit travel times are computed with a purpose-built time-dependent search over the MTA static GTFS feed (implemented in Python; no external routing engine), from each listing centroid to each of three workplace anchors — Midtown Manhattan, Downtown Manhattan/Financial District, and Downtown Brooklyn — under an arrive-by constraint of 09:00 on a representative weekday. Approximately 4,000 × 3 = 12,000 routings.

Departure time is fixed rather than varied because no hypothesis concerns time-of-day variation, and three anchors are used rather than the larger set initially contemplated for the same reason. Walk access to and from stops is computed as straight-line distance with a 1.3 detour factor at 4.8 km/h rather than routed over the street network, within a **1,200 m** access radius. The radius was set empirically rather than by convention. At the literature-standard 800 m, 11% of listings fall outside walking range of any station, and those listings are **$600/month cheaper at the median** and concentrated in Queens (28.7% of the excluded versus 13.2% of the retained), Staten Island (14.0% versus 1.1%) and the Bronx (19.8% versus 11.0%) — that is, the filter removes precisely the cheap, bus-dependent outer-borough inventory that a study of rent gaps cannot afford to lose. Widening to 1,200 m retains 94.6% of listings while moving the median commute only from 30.0 to 30.4 minutes. The 800 m universe is reported as a robustness check. Buses remain excluded; this is the residual cost of that exclusion, and it is measured rather than assumed.

The Staten Island Railway has no track connection to the subway, so a subway-only network renders every Staten Island listing unreachable and silently deletes a borough. We add the Staten Island Ferry explicitly as a fixed-cost link (25 minute crossing plus 15 minute mean wait on a 30 minute headway).

### *5.3 Tract covariates*

ACS 5-year estimates at census-tract level were specified as: median household income, median gross rent, rent burden, renter share, and racial and ethnic composition. They were to enter the analysis **only** as the neighborhood-exposure outcome (§6, S3) and as descriptive context, and are not inputs to the feasible set, the dominance computation, or any primary outcome.

These covariates **were** retrieved, in `code/17_fetch_acs.py`, for all **2,327** census tracts in the five New York counties, matching the tract count in the TIGER geometry used for the spatial join. 99.5% of listings resolve to a tract with a non-suppressed median-income estimate; the remainder fall in tracts where ACS suppresses the estimate for small population, and are dropped from S3 only.

An earlier version of this paper reported these covariates as not retrieved and dropped the S3 measure accordingly. That was a consequence of the Census API key not yet being available, not of any design decision, and it has been corrected: S3 is reported in §8.11.

### *5.4 Coverage benchmark*

New York's rental market transacts substantially through StreetEasy and REBNY channels rather than MLS syndication, so a syndication-derived sample may under-represent no-fee units and small-landlord inventory. We benchmark the listing sample's rent distribution and borough composition against ACS median gross rent by tract and against the NYC Housing and Vacancy Survey, and report the direction and magnitude of divergence in **Table 2b**; Table 2a reports a separate and fully computed exclusion check, the walk-access radius. This does not eliminate the bias; it makes it legible, and it bounds the claims in Section 9.

**Table 2a. Direction of the walk-access exclusion (n = 4,108 geocoded listings).**

| | **Retained (n = 3,885)** | **Excluded (n = 223)** |
|---|---|---|
| Median rent | $4,200 | **$3,500** |
| Mean rent | $4,783 | $3,681 |
| Median bedrooms | 1 | 2 |
| Manhattan share | 37.3% | 1.3% |
| Brooklyn share | 36.0% | 23.8% |
| Queens share | 13.5% | **39.5%** |
| Bronx share | 11.7% | 16.1% |
| Staten Island share | 1.6% | **19.3%** |

Excluded listings are $700/month cheaper at the median and overwhelmingly outer-borough. At the 800 m radius the exclusion was larger (449 listings, $600/month cheaper); the 1,200 m radius adopted here halves it. The residual bias is toward *over-representing* expensive, subway-proximate inventory, which means our absolute rent levels are high relative to the true market and our cost-gap magnitudes should be read as pertaining to the transit-accessible segment.

**Table 2b. Rent and borough distribution of the listing sample against NYCHVS 2023 and ACS 2023.** Gross rent, weighted by the NYCHVS final household weight `FW`. The NYCHVS rent benchmark requires no API key; the ACS borough benchmark below requires one and it was obtained (Appendix D, item 9). Tenure coding was verified rather than assumed — the renter code carries positive gross rent and zero owner cost, and yields a weighted renter share of **67.7%**, matching the published New York figure. Cases coded no-cash-rent are excluded, so the benchmark is the distribution of cash gross rent.

| Decile | Sample (asking rent) | NYCHVS: all renters | NYCHVS: moved 2021–23 | NYCHVS: moved 2021–23, unsubsidised |
|---|---|---|---|---|
| p10 | $2,867 | $613 | $1,150 | $1,283 |
| p20 | $3,224 | $1,031 | $1,543 | $1,598 |
| p30 | $3,500 | $1,270 | $1,765 | $1,850 |
| p40 | $3,850 | $1,484 | $2,019 | $2,106 |
| p50 | **$4,200** | **$1,694** | **$2,300** | **$2,385** |
| p60 | $4,575 | $1,924 | $2,700 | $2,810 |
| p70 | $5,250 | $2,201 | $3,123 | $3,221 |
| p80 | $6,103 | $2,660 | $3,608 | $3,708 |
| p90 | $7,680 | $3,470 | $4,391 | $4,493 |
| *n* | 3,885 listings | 5,893 records | 1,180 | 1,038 |

**The comparison must be read as flow against stock, not sample against population.** Our sample is *asking* rent on units currently available. NYCHVS gross rent over all renters is rent currently *paid*, and New York's occupied stock contains a large body of rent-stabilised, rent-controlled and subsidised tenancies of long duration that no listings sample can contain at any sampling intensity. Benchmarking against all renters therefore attributes a structural feature of the housing stock to our sampling, and overstates the coverage problem. We report three nested populations so the two effects can be separated, and treat **recent unsubsidised movers** as the fair comparison, since that is the market-rate flow a renter using a listings site actually faces.

| Benchmark population | Sample median as a percentile | Ratio |
|---|---|---|
| All renter households, cash rent | 95.0th | 2.48× |
| Moved in 2021–23 | 88.2th | 1.83× |
| Moved in 2021–23, no rent assistance | **87.0th** | **1.76×** |

**The honest statement is 1.76×, not 2.48×.** Against the market-rate flow, the median listing in our sample rents for 1.76 times the median recent unsubsidised letting, and sits at roughly the **87th percentile** of that distribution. The sample is drawn from approximately the upper eighth of the New York rental market by price. This is a substantial upward bias and it is consistent in direction with both the walk-access exclusion (Table 2a) and the borough composition above.

Three consequences, stated in the terms Section 9 needs:

1. **The internal comparisons are unaffected.** Every primary outcome is computed within a scenario against a pool held fixed across conditions, so a level shift in the rent distribution cannot move a dominance rate or a within-scenario contrast. Nothing in §8 depends on the sample being representative.
2. **The dollar magnitudes are segment-specific.** The $900/month median dominance gap and the +$606/month residual gap pertain to the transit-accessible, market-rate, upper-decile segment. They should not be read as the loss facing a median New York renter, and we do not claim they are. Whether the gap scales with rent level, is constant in dollars, or is roughly proportional is not identified by our design.
3. **The direction of the likely error is knowable.** If the gap is proportional to rent, our absolute dollar figures overstate the loss for a median renter while the *relative* loss travels; if it is roughly constant in dollars, they transfer directly. A sample spanning the lower deciles would settle this and is the single highest-value extension to the data collection.

**Borough composition against ACS renter-occupied units.** ACS 2023 5-year, table B25003, all five New York counties. NYCHVS borough shares are shown alongside as an independent check; the two sources agree to within 1.2 percentage points on every borough, which is reassuring for both.

| Borough | Sample share | ACS renter-occupied | NYCHVS renter households | Sample ÷ ACS |
|---|---|---|---|---|
| Manhattan | 37.3% | 26.2% | 25.3% | **1.42×** |
| Brooklyn | 36.0% | 31.9% | 30.9% | 1.13× |
| Queens | 13.5% | 20.5% | 20.8% | **0.66×** |
| Bronx | 11.7% | 19.0% | 20.1% | **0.62×** |
| Staten Island | 1.6% | 2.4% | 2.9% | 0.67× |
| *n* | 3,885 listings | 2,226,896 units | 2,026,023 households | — |

Manhattan is over-represented by roughly half again, and the Bronx, Queens and Staten Island are each under-represented by a third or more. The composition bias therefore runs in the same direction as the rent bias and compounds it: the sample is drawn disproportionately from the most expensive borough, and within that borough from the upper deciles. Table 2b is now complete.

### *5.5 Exploratory data analysis plan*

EDA is specified in advance and its outputs are descriptive; no primary hypothesis is formed from it.

**Table 3a. Cleaning cascade.**

| n | Step |
|---|---|
| 4,257 | Raw records retrieved (10 API requests) |
| 4,257 | After deduplication on listing id |
| 4,257 | Status = Active |
| 4,257 | Has coordinates |
| 4,190 | Has bedroom count |
| 4,190 | Has rent |
| 4,108 | Rent within p1–p99 ($1,600–$13,533) |
| 4,108 | After deduplication on address \| bedrooms \| rent |
| 4,108 | Assigned to an NYC census tract (0 fell outside) |
| **3,885** | **Reachable by subway within a 1,200 m walk (223 excluded)** |

Queens (649 records) and Staten Island (108) were retrieved to *exhaustion* and are therefore complete rather than sampled; Manhattan and Brooklyn are truncated at 1,500 each under the API's default ordering (see §5.1 sampling caveat). Floor area is missing for 39% of records and is not used as a constraint.

**Table 3b. Median rent by bedrooms and borough (USD/month).**

| Bedrooms | Bronx | Brooklyn | Manhattan | Queens | Staten Is. |
|---|---|---|---|---|---|
| Studio | 2,548 | 3,530 | 3,954 | 3,156 | 2,878 |
| 1 BR | 2,995 | 4,095 | 4,950 | 3,446 | 1,900 |
| 2 BR | 3,300 | 4,625 | 6,899 | 4,780 | 2,700 |
| 3 BR | 3,628 | 4,195 | 7,495 | 4,244 | 3,366 |

**Table 3c. Listing counts by bedrooms and borough.**

| Bedrooms | Bronx | Brooklyn | Manhattan | Queens | Staten Is. | Total |
|---|---|---|---|---|---|---|
| Studio | 32 | 186 | 238 | 86 | 4 | 546 |
| 1 BR | 191 | 567 | 605 | 248 | 9 | 1,620 |
| 2 BR | 121 | 430 | 371 | 140 | 19 | 1,081 |
| 3 BR | 86 | 189 | 183 | 39 | 26 | 523 |
| 4+ BR | 24 | 27 | 51 | 10 | 3 | 115 |

**Table 4. Transit commute to workplace anchors (minutes, arrive by 09:00; n = 3,885).**

| | Midtown | Downtown Manhattan | Downtown Brooklyn |
|---|---|---|---|
| Mean | 33.8 | 33.9 | 35.3 |
| SD | 16.4 | 15.1 | 17.3 |
| p10 | 16.7 | 16.5 | 13.9 |
| p25 | 21.2 | 25.1 | 23.0 |
| Median | 30.4 | 32.1 | 34.3 |
| p75 | 42.6 | 40.6 | 43.5 |
| p90 | 53.4 | 52.4 | 56.8 |
| Max | 118.9 | 103.2 | 103.2 |
| Within 30 min | 49.0% | 42.9% | 37.7% |
| Within 45 min | 80.6% | 82.5% | 77.2% |
| Within 60 min | 93.4% | 93.4% | 91.5% |
| **Corr. with rent** | **−0.415** | **−0.362** | **−0.293** |

Router validation against known routes (to Midtown): Grand Central 10.2 min, Wall St 21.9, Bedford Av 21.9, Jay St–MetroTech 27.9, Flushing–Main St 34.5, Coney Island 61.0, Far Rockaway 86.5, St George 66.0 (via ferry), Tottenville 106.5. All within the range of published travel times.

The rent–commute correlations are the material result in this table. They are negative, as theory requires, but far from −1. Had they approached −1, rent and commute would collapse onto a single dimension and strict dominance would lose all discriminating power (design check B, §5.5).

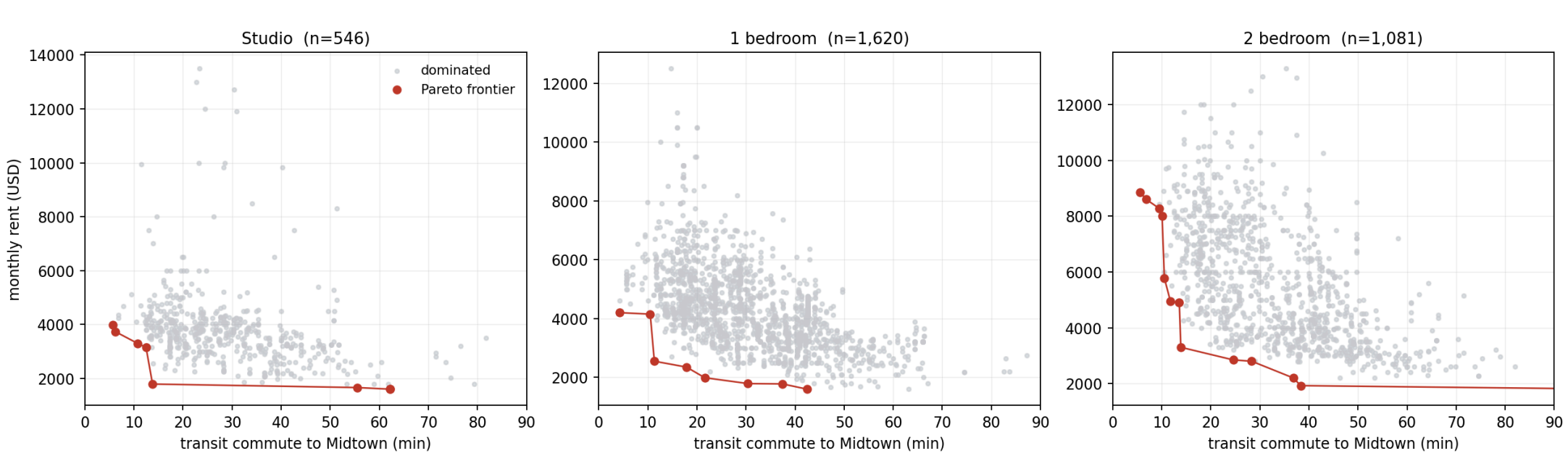

Figure 1

**Figure 1. Rent–commute structure of the listing universe, with Pareto frontiers, by bedroom count.** Each panel plots all listings of one bedroom type on transit commute to Midtown against monthly rent. Highlighted points are the Pareto frontier: no listing in the sample is both cheaper and faster. Frontier sizes in the full universe are 8 of 546 studios (1.5%), 8 of 1,620 one-bedrooms (0.5%), and 13 of 1,081

two-bedrooms (1.2%). The scale of the dominated region is the visual statement of the paper's construct — and the reason P2 must be reported separately against the pool and against the universe (§6.1).

**Table 5. Feasible-set descriptives across the 150 scenarios.**

| | Mean | p5 | p25 | Median | p75 | p95 | Max |
|---|---|---|---|---|---|---|---|
| \|F_i\| in universe | 605.8 | 77 | 288 | **511** | 830 | 1,499 | 1,881 |
| Pareto frontier, universe | 18.1 | 6.4 | 12 | 17 | 23 | 30.5 | 34 |
| \|F_i\| in pool | 40.0 | 40 | 40 | 40 | 40 | 40 | 40 |
| Pareto frontier, pool | 18.1 | 6.4 | 12 | **17** | 23 | 30.5 | 34 |
| Budget (USD) | 4,227 | 3,100 | 3,600 | 4,100 | 5,000 | 6,950 | 6,950 |
| Min feasible rent | 1,700 | 1,600 | 1,600 | 1,600 | 1,625 | 2,700 | 2,800 |
| Min feasible commute (min) | 6.1 | 2.2 | 2.7 | 4.0 | 10.5 | 11.8 | 13.9 |

Frontier size is identical in the universe and in the pool because pools are constructed to contain the entire universe frontier (§4.4); the largest observed frontier, 34, fits within the 40 feasible slots.

**Design check outcomes.**

- **Check A** (median |F_i| ≥ 10): median 511, with **0 of 150** scenarios below 10. **Pass.** The budget grid is correctly calibrated and no redraw is required.
- **Check B** (median pool Pareto frontier > 2): median **17** of 40 feasible listings. **Pass.** Approximately 55% of feasible pool listings are dominated, giving the primary metric substantial room to discriminate without being degenerate in either direction.

Feasible-set size behaves monotonically in the design factors, as it should: median |F_i| rises from 287 at the 25th budget percentile to 862 at the 75th, and from 277 at a 30-minute commute ceiling to 713 at 60 minutes.

Two EDA results are pre-specified as **design checks with stopping implications**. If the median |F_i| is below 10, the scenario budget grid is mis-calibrated and must be re-drawn before any model is called. If rent and commute are so strongly correlated that Pareto frontiers are near-degenerate (median frontier size ≤ 2), the dominance metric loses discriminating power and we must either widen the pool or add a third dominance dimension; this contingency and its trigger are registered in advance.

---

## 6. Measures

Primary outcomes are **weight-free**: none depends on a researcher-chosen utility function. This is a deliberate defense against the standard objection that opportunity-loss findings merely restate the analyst's preferences.

Notation: for scenario *i*, let P_i be the candidate pool, F_i ⊆ P_i the feasible set, and R_i the set of five recommended listings.

**Feasible set.**

```
F_i = { j ∈ P_i : Rent_j ≤ Budget_i
                AND Bedrooms_j ≥ Bedrooms_i
                AND Commute_ij ≤ MaxCommute_i }
```

**Strict dominance.** Listing j is strictly dominated by j′ ∈ P_i iff

```
Rent_j′     ≤ Rent_j
Commute_ij′ ≤ Commute_ij
Bedrooms_j′ ≥ Bedrooms_j
```

with at least one inequality strict. The Pareto frontier of F_i is the set of undominated feasible listings.

### *6.1 Primary outcomes (pre-specified)*

**P1 — Confirmed hard-constraint violation rate.**

```
ViolationRate_i = |{ j ∈ R_i : j violates a verified constraint }| / |R_i|
```

Reported overall and decomposed into over-budget, under-bedroom, and over-commute. Listings whose violation status depends on an `unknown` field are counted as *unverifiable* and excluded from the numerator, so P1 is a lower bound.

**P2 — Strict-dominance rate.** Reported as two distinct quantities, because the measure is sensitive to the size of the set it is computed against. In our data the full universe contains 1,539 one-bedroom listings of which only 8 are undominated (99.5% dominated), whereas a 40-listing feasible pool contains roughly 17 undominated members (~55% dominated). A single "dominance rate" would therefore conflate model behavior with our own sampling parameter.

*P2a — Pool dominance (primary).* Dominated by a listing the model actually saw:

```
PoolDominance_i = |{ j ∈ R_i : ∃ j′ ∈ F_i ⊆ P_i, j′ strictly dominates j }| /
|R_i|
```

This is the accountability measure: the system was shown 120 listings and returned one beaten, on the stated criteria, by another listing on the same screen. Pool size is exactly 120 for all 150 scenarios and identical across the four identity conditions within a scenario, so identity contrasts are unaffected by the sensitivity. The $N \in \{60, 120\}$ robustness arm is **mandatory** rather than optional, and exists specifically to demonstrate that the primary estimate is not a pool-size artifact.

*P2b — Universe dominance (secondary; specified, not computed in this version).* Dominated by any listing in the full inventory, whether retrieved or not. It would be labelled explicitly as a **retrieval-plus-ranking composite**, not a model failure, since our design fixes retrieval by construction. Because our pools are built to contain the entire universe Pareto frontier, P2a is a conservative lower bound on what a user faces in a system that must also retrieve. No P2b figure is reported in Section 8.

Both are reported with the *magnitude* of dominance: for each dominated recommendation, the rent difference in dollars per month and the commute difference in minutes to its dominating listing, summarized by median and interquartile range.

**P3 — Cost gaps.**

```
RentGap_i    = median{ Rent_j : j ∈ R_i ∩ F_i }     − median{ Rent_j : j ∈ OracleTop5_i }
CommuteGap_i = median{ Commute_ij : j ∈ R_i ∩ F_i } − median{ Commute_ij : j ∈ OracleTop5_i }
```

Units are USD per month and minutes. The intersection with F_i is essential: taking a statistic over R_i unrestricted allows a high-scoring infeasible listing to mask the loss precisely in the cases the measure exists to detect. Medians rather than minima are used because a minimum is determined by a single listing and is unstable under ties.

The identity contrast is the within-scenario difference in these gaps across conditions.

### *6.2 Secondary outcomes*

**S1 — Mean percentile rank within the feasible set.** The mean percentile of recommended listings within F_i, computed separately on rent and on commute. This replaces top-*k* capture as the primary opportunity measure: when |F_i| is large and members are near-equivalent, Capture@5 approaches zero for any system including an optimal one, because it measures set size rather than recommendation quality.

**S2 — Refusal and information-withholding rate.** The share of responses that decline to recommend, decline to discuss neighborhood characteristics, or substitute a safety or fair-housing statement for substantive content. **As executed, this was detected by a keyword pattern** matching refusal and fair-housing hedging language over the first 600 characters of each response, not by human coding. Because the observed rate was exactly 0.0% across all 5,916 parsed responses (Table 11) and every response contained five valid listing ids, we did not proceed to the planned human double-coding: there were no candidate cases to adjudicate. A non-zero rate would require the rubric-based protocol originally specified.

**S3 — Neighborhood exposure.** Distribution of ACS tract characteristics across recommended listings: tract median household income, median gross rent, rent burden, renter share, and racial and ethnic composition. For each response we take the recommended listings, resolve each to its census tract, and summarise the set by the median tract value. The identity contrast is the within-scenario difference against the neutral baseline, using the same estimand and the same randomization procedure as every other outcome (§7.2). This is the measure that connects our results to the steering literature and permits comparison with Liu et al. (2024) and Samad et al. (2026), subject to the design limit stated in §8.11. Reported in §8.11 (Tables 22–24).

**S4 — Conventional IR metrics.** Tolerance-band Capture@k (a recommendation counts as a hit if within $50 and 5 minutes of an oracle top-5 member), NDCG@5, and Precision@5. All relevance labels derive from the pre-specified constraints and benchmark; none is generated post hoc or by a model judging itself.

### *6.3 Tertiary (robustness only)*

**T1 — Weighted opportunity loss.**

```
Score_ij = −w1·Rent_j − w2·Commute_ij + w3·UnitMatch_ij + w4·TransitAccess_j

OpportunityLoss_i = max_{j ∈ F_i} Score_ij − max_{j ∈ R_i ∩ F_i} Score_ij
```

with weights derived from the scenario's stated priority ordering, reported across a sensitivity grid of at least three weight vectors. **No headline claim rests on T1.** It is included to connect with a literature that expects a scalar utility measure, and to demonstrate that our conclusions do not depend on one.

### *6.4 Fairness gap*

For any outcome Y,

```
FairnessGap(A, B) = E[ Y_i | Identity = A ] − E[ Y_i | Identity = B ]
```

estimated as a within-scenario matched contrast, since scenario, pool, and pool-ordering seed are held constant across conditions by construction.

---

## 7. Statistical Analysis

### *7.1 Design*

The study is a **within-scenario matched design**: the four identity conditions face an identical request and an identical candidate pool. Identity effects are matched contrasts, not between-group comparisons, and the effective sample size for identity inference is the number of scenarios (150), not the number of model calls (6,840). We state this explicitly because treating calls as independent observations is the characteristic inferential error of large-scale LLM audits.

### *7.2 Primary inference: randomization*

Because condition assignment is randomized within scenario, randomization inference is the method matched to the design. We permute identity labels within scenario across 10,000 draws to construct the null distribution of each primary contrast, and report the observed paired difference with a permutation *p*-value and a percentile confidence interval.

This is preferred to cluster-robust OLS for a substantive reason: P1 and P2 are zero-inflated proportions, and with 150 clusters the asymptotic properties of clustered standard errors are unreliable in exactly the regime where our effects are expected to lie.

### *7.3 Secondary inference: fixed-effects regression (specified, not estimated)*

The models below were pre-specified as secondary analyses. **They were not estimated for this version.** Because the primary randomization-inference contrasts returned no effect surviving correction (§8.4), and because the variance decomposition showed the condition component to be smaller than replicate noise (Table 10), fitting a fixed-effects model to recover the same null would add no information. They are retained here so the specification is on record and so any revision reporting them cannot be accused of having chosen the model after seeing the data.

*Model 1 — identity disparities.*

```
Y_igmr = α_i + β·IdentityCue_g + λ_m + δ_r + ε_igmr
```

*Model 2 — mitigation and interaction.*

```
Y_igsmr = α_i + β1·Grounded_s + β2·ConstraintFirst_s + β3·IdentityCue_g
          + β4·(ConstraintFirst_s × IdentityCue_g) + λ_m + δ_r + ε_igsmr
```

where *i* indexes scenario, *g* identity condition, *s* architecture, *m* model, *r* replicate. β4 would be the coefficient of record for RQ4. Standard errors clustered at scenario level.

### *7.4 Variance decomposition*

A mixed model partitions variance into between-scenario, between-condition, between-name (within condition), and within-cell replicate components. **A pre-specified interpretive rule:** if the replicate variance component exceeds the identity component, the identity effect is not reported as systematic bias regardless of its *p*-value. This rule exists because temperature-1.0 sampling can manufacture apparent disparities that are indistinguishable from noise, and the audit literature has not consistently guarded against it.

### *7.5 Power and achieved precision*

The planned simulation-based power analysis on a 20-scenario pilot **was not run**; the design proceeded directly to the full grid because the marginal cost of doing so was under US$6. We therefore report *achieved precision* rather than planned power.

With 150 matched scenarios and three replicates per cell, the observed 95% percentile intervals on the paired identity contrasts (Table 9) span roughly ±$350 on rent gap and ±0.15 on dominance rate. This is the honest statement of what the null in §8.4 excludes: identity effects larger than approximately $350/month on rent gap would very likely have been detected; effects of, say, $50/month would not. The null is well-powered against *large* disparities and uninformative about small ones. Readers should not interpret it as evidence of exact equality.

### *7.6 Safeguards*

- **Pre-registration — planned but NOT lodged.** The protocol in §§3–7 was written and version-controlled before any main-grid call, and the design-check stopping rules of §5.5 were evaluated before model calls began. However, **no pre-registration was filed with OSF or any registry.** We state this plainly rather than describe the intent as if it were the act. The consequence is that readers must take the analysis plan's priority on the strength of the timestamped repository history rather than on an independent registry record, and any future version of this work should register before collecting additional data. The primary outcomes, the §7.4 interpretive rule, and the family definitions for multiple-testing correction were all fixed in the v0.1 manuscript dated 23 August 2026, prior to execution.
- **Multiple testing**: Benjamini–Hochberg correction within pre-declared outcome families (constraint fidelity; opportunity; identity; withholding). Family membership is fixed at registration.
- **Randomization** of pool order per scenario, held identical across that scenario's identity conditions; **rotation** of names across scenarios. Prompt-wording variants were specified but not executed (§4.5).
- **Snapshot logging**: model identifier, API version, call timestamp, and full request/response recorded for every call.

- **Deviation log**: any departure from the registered protocol recorded with date and rationale, reported in Appendix C.

---

## 8. Results

We report **6,840 attempted calls** across three models and two vendors:

| Model | Vendor | Design | Attempted | Parsed |
|---|---|---|---|---|
| `gpt-5.6-luna` | OpenAI | 150 scen × 4 cond × 3 arch × 3 reps | 5,400 | 5,218 (96.6%) |
| `gpt-5.6-sol` | OpenAI | 60 scen × 4 cond × S1 × 3 reps | 720 | 698 (97.0%) |
| `claude-opus-5` | Anthropic | 60 scen × 4 cond × S1 × 3 reps | 720 | 715 (99.3%) |
| **Total** | | | **6,840** | **6,631 (97.0%)** |

**Zero hallucinated listing ids** were observed across all parsed responses in any arm.

**Table 12b. Complete call and cost reconciliation across every arm reported in this paper.**

| Arm | Attempted | Parsed | Cost |
|---|---|---|---|
| Main grid (§8.1–8.7) | 6,840 | 6,631 (96.9%) | $47.61 |
| Priority manipulation, luna (§8.8) | 1,350 | 1,338 (99.1%) | $1.27 |
| Priority manipulation, Claude (§8.8) | 165 | 157 (95.2%) | $6.57 |
| Pool density (§8.9) | 800 | 784 (98.0%) | $0.81 |
| Size sweep (§8.10) | 790 | 691 (87.5%) | $0.75 |
| **Total** | **9,945** | **9,601 (96.5%)** | **$57.01** |

Sections 8.1–8.7 report the main grid only; §8.8–8.10 report the robustness and diagnostic arms, each with its own n stated in place. Costs are $18.46 on OpenAI and $38.55 on Anthropic. The size-sweep arm was purged of 555 records created by a spend-ledger race condition (Appendix C, deviation 8); those attempts never reached the provider and cost nothing. Parse failures are analyzed for treatment-correlated missingness in §8.7.

### *8.1 Constraint fidelity (RQ1)*

**Table 6. Confirmed hard-constraint violation rate, `gpt-5.6-luna`.**

| Architecture | n | Violation | Over budget | Under bedrooms | Over commute |
|---|---|---|---|---|---|
| S1 direct | 1,743 | 1.76% | 0.08% | 0.03% | 1.64% |
| S2 grounded | 1,738 | 1.64% | 0.05% | 0.01% | 1.58% |
| S3 constraint-first | 1,737 | **0.00%** | 0.00% | 0.00% | 0.00% |
| *S_rand chance floor* | — | *66.6%* | — | — | — |

**H1a is rejected for these models.** Compliance is near-perfect: with 80 of 120 pool listings violating exactly one constraint, a system selecting at random would violate on 66.6% of picks; direct prompting violates on 1.8%. Budget — the constraint most legible in the listing record — is violated on 0.08% of recommendations, essentially never. What residual violation exists is almost entirely commute (93% of all violations), the one constraint requiring the model to read a supplied numeric field rather than compare two prices.

S3's zero rate is definitional, not empirical, and is reported only as an implementation check (§3, H4a).

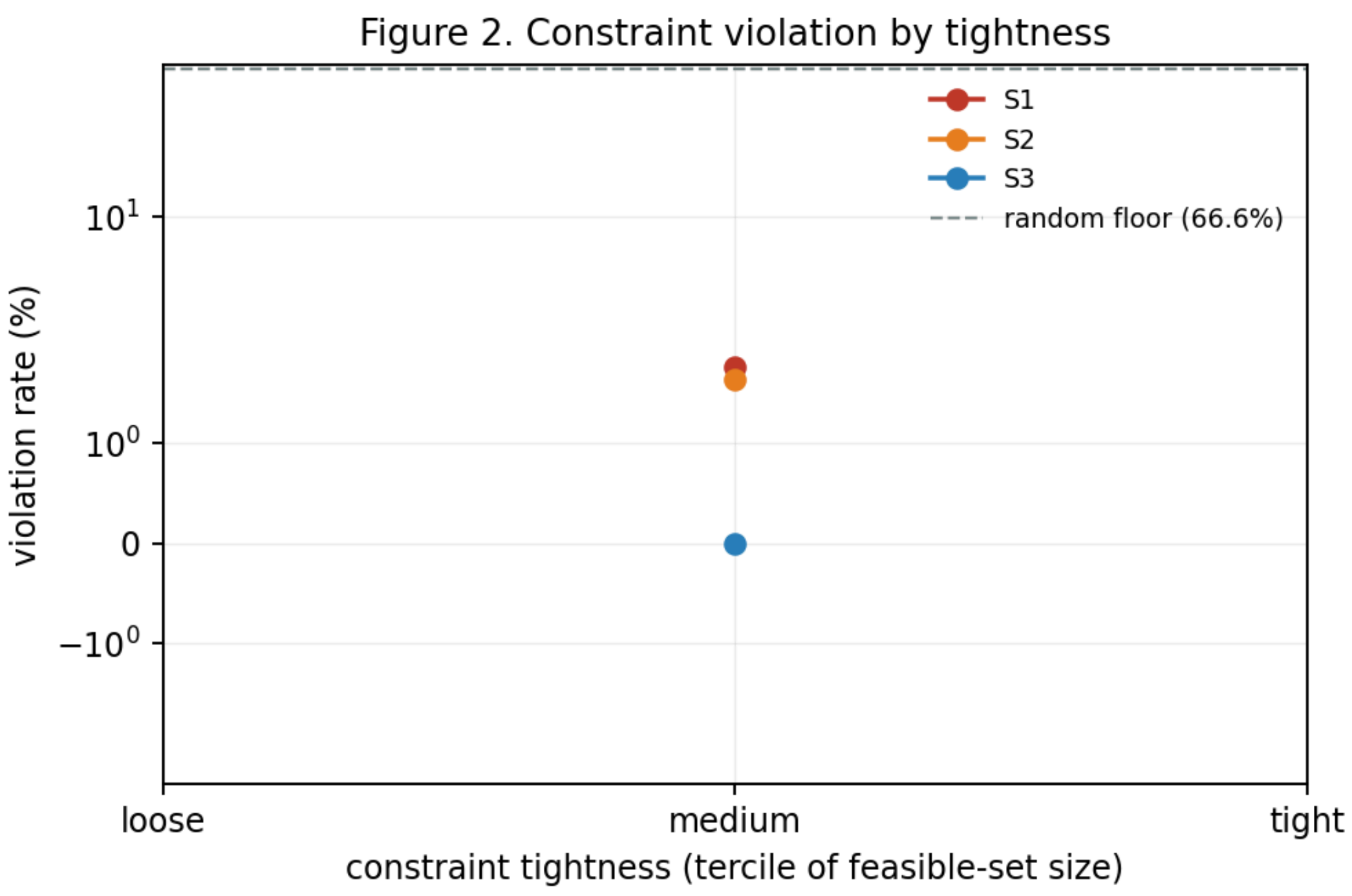

Figure 2

**Figure 2. Constraint violation by constraint tightness.** Violation rate (%) against terciles of feasible-set size, by architecture, with the random-selection floor at 66.6%. The vertical axis is symmetric-log because S3's rate is identically zero by construction. All three architectures sit one to two orders of magnitude below chance across every tightness tercile, and the ordering S3 < S2 < S1 is stable. Tightness moves the rate only slightly, so near-perfect compliance is not an artifact of loose constraints: it holds where the feasible set is smallest and a careless selector would be most likely to fail.

### *8.2 Forgone opportunity (RQ2)*

**Table 7. Opportunity measures, `gpt-5.6-luna`.**

| Architecture | n | Dominance rate | Rent gap | Commute gap | Mean rent percentile in F_i |
|---|---|---|---|---|---|
| S1 direct | 1,743 | 39.0% | **+$498/mo** | −4.0 min | 0.355 |
| S2 grounded | 1,738 | 39.1% | +$506/mo | −4.0 min | 0.362 |
| S3 constraint-first | 1,737 | 34.8% | +$467/mo | −4.0 min | 0.355 |
| *S_rand chance floor* | — | *55.1%* | *+$261/mo* | *+10.2 min* | — |

**H2a is supported.** Roughly two in five recommendations are strictly dominated — beaten on rent *and* commute *and* bedroom count by a listing on the same screen. The model is meaningfully better than chance at avoiding dominated listings (39.0% vs 55.1%), so it is not selecting arbitrarily. But **on price specifically it underperforms random selection**, with a rent gap nearly double the chance floor. These two facts are compatible and jointly diagnostic: the model avoids listings that are bad on *every* dimension while systematically overpaying, because it is buying proximity (commute gap −4.0 min against a +10.2 min floor).

**Table 7b. Magnitude of dominance — gap to the best dominating listing (12,281 dominated recommendations across all three models).**

| Model | Arch | n | Median $/mo | Mean $/mo | p75 $/mo | Median extra minutes |
|---|---|---|---|---|---|---|
| luna | S1 | 3,346 | **900** | 945 | 1,275 | 3.5 |
| luna | S2 | 3,334 | 900 | 934 | 1,310 | 3.4 |
| luna | S3 | 3,022 | 900 | 949 | 1,310 | 3.4 |
| sol | S1 | 1,364 | 960 | 1,035 | 1,395 | 3.7 |
| claude-opus-5 | S1 | 1,215 | 900 | 871 | 1,250 | 3.5 |
| *OpenAI models only* | — | *11,066* | *900* | *949* | *1,310* | *3.5* |

When a recommendation is dominated, the listing that beats it is typically **$900/month cheaper and 3.5 minutes closer** — simultaneously. This is the paper's central quantity, and it requires no assumption about what the user values.

**Architecture barely helps.** Constraint-first reduces dominance from 39.0% to 34.8% and rent gap from $498 to $467 — real but modest. Prompt-level grounding (S2) does nothing at all (39.1%, +$506). **The**

**mitigation that eliminates every hard-constraint violation leaves ~89% of the opportunity loss intact**, which localizes the failure firmly in the soft ranking stage rather than in constraint comprehension.

Figure 3. Priced opportunity loss

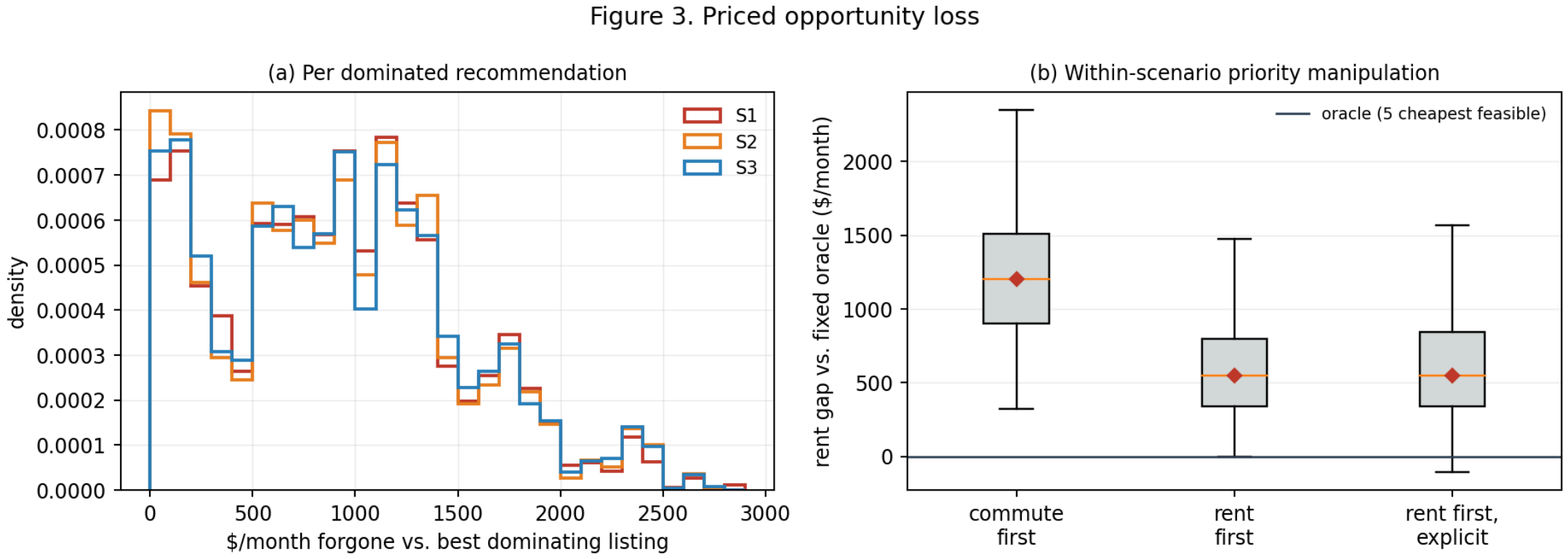

Figure 3

**Figure 3. Priced opportunity loss.** (a) Distribution of the rent gap to the best dominating listing, per dominated recommendation, by architecture (`gpt-5.6-luna`). The mass well to the right of zero is the paper's central quantity: it is not a near-miss distribution. Architecture shifts it barely. (b) The within-scenario priority manipulation of §8.8, plotted against the *single fixed* oracle used for every condition — the five cheapest feasible listings in that scenario's pool. Two things are visible at once. The drop from commute-first to rent-first (median $1,200 to $550) is responsiveness: the model moves in the correct direction when the stated priority changes. But every box sits far above the oracle line, and adding an unambiguous lexicographic instruction ("rent first, explicit") moves the median not at all. Responsiveness and optimization are separate capabilities, and only the first is present.

**Note on panel (b).** An earlier version of this figure grouped scenarios by their *own* stated priority against a priority-defined oracle — the between-scenario comparison reported in Table 8 below and retracted in §8.3. Because the scenarios and the oracle both move with the condition, that version could not separate responsiveness from selection, and it is not shown. Panel (b) holds the scenario, the pool, the candidate ordering and the oracle fixed and varies one sentence.

### *8.3 Opportunity loss varies with the stated priority (superseded by §8.8)*

**Table 8. Rent gap by the user's stated priority (S1 direct, `gpt-5.6-luna`, between-scenario).**

| User's stated priority | n | Mean rent gap | Median | Commute gap | Dominance |
|---|---|---|---|---|---|
| "Rent matters most" | 586 | +$624/mo | +$550 | −4.2 min | 39% |
| "Location matters most" | 582 | +$1,122/mo | +$1,050 | −11.7 min | 40% |
| "Commute matters most" | 575 | −$260/mo | −$105 | +4.1 min | 31% |

**This table is confounded and we report it only for transparency.** Priority is a *between-scenario* factor in the main grid: each scenario carries exactly one priority, so these rows compare different scenarios with different budgets, workplaces and commute ceilings. Worse, the oracle is *defined by* the stated priority (§6.1), so each row is scored against a different benchmark.

Our first reading of this table was that models fail to honor stated preference orderings — a "preference infidelity" claim. **That reading was wrong.** §8.8 reports a within-scenario manipulation that isolates the effect properly and reaches the opposite conclusion on responsiveness, while sustaining and sharpening the conclusion on optimization. Readers should treat §8.8, not Table 8, as the finding.

### *8.4 Identity-conditioned disparity (RQ3) — no evidence of large effects*

**Table 9. Within-scenario identity contrasts, randomization inference (10,000 permutations, Benjamini–Hochberg within outcome family).**

**Zero of 36 contrasts are significant after correction.** Representative results (S1, n=150 paired scenarios):

| Metric | Contrast | Observed | p | p (BH) |
|---|---|---|---|---|
| Dominance | C1 name A − C0 | +0.003 | 0.677 | 1.000 |
| Dominance | C3 voucher − C0 | −0.001 | 0.940 | 0.940 |
| Rent gap | C1 name A − C0 | +$5.32 | 0.694 | 1.000 |
| Rent gap | C2 name B − C0 | +$7.82 | 0.532 | 0.958 |
| Rent gap | C3 voucher − C0 | −$17.48 | 0.247 | 0.555 |
| Commute gap | C3 voucher − C0 | +0.28 min | 0.110 | 0.993 |
| Violation | C1 name A − C0 | +0.005 | 0.080 | 0.724 |

**Table 10. Variance decomposition (S1, `gpt-5.6-luna`).**

| Outcome | Between-scenario | Between-condition | Within-cell (replicate) | Verdict |
|---|---|---|---|---|
| Dominance | 0.06 | 0.0000 | 0.01 | condition < replicate noise |
| Rent gap | 563,122 | 164 | 36,132 | condition < replicate noise |

Under the interpretive rule pre-specified in §7.4 — if the replicate variance component exceeds the condition component, the effect is not reported as systematic bias regardless of *p*-value — **we report no identity-conditioned disparity.** The variability produced by asking the same model the same question twice exceeds the variability produced by changing the identity cue.

**Table 11. Refusal and information withholding.**

| Condition | n | Refusal / withholding rate |
| --- | --- | --- |
| C0 neutral | 1,286 | 0.0% |
| C1 name A | 1,306 | 0.0% |
| C2 name B | 1,318 | 0.0% |
| C3 voucher | 1,308 | 0.0% |

**H4b is rejected.** No refusals, no substitution of safety language for substance, no differential withholding. The guardrail trade-off we anticipated did not materialize in this setting.

**What this null does and does not license.** It is a well-powered negative result for *ranking over a fixed candidate pool*: 150 matched scenarios, byte-identical requests and pools across conditions, 10,000-draw randomization inference. It is **not** evidence that these models do not steer. Prior work finding steering (Liu et al., 2024; Samad et al., 2026) used open-ended prompts in which the model selects neighborhoods itself. Our design removes that degree of freedom by construction. Whether fixing the candidate set *causes* the disparity to vanish is a plausible hypothesis that **this study does not test**, because we did not run an open-ended comparison arm. Section 9.4 specifies that experiment.

### *8.5 Capability does not reduce opportunity loss*

**Table 12. `gpt-5.6-luna` vs `gpt-5.6-sol`, S1 only, 60 common scenarios.**

| Model | n | Violation | Dominance | Rent gap | rent-first gap |
| --- | --- | --- | --- | --- | --- |
| gpt-5.6-luna | 706 | 1.7% | 37.6% | +$621 | +$700 |
| gpt-5.6-sol (17× cost/token) | 698 | 1.5% | 39.4% | +$649 | +$699 |

The flagship model is statistically indistinguishable from the budget tier on every primary outcome, and the rent-first gap is essentially identical (+$699 vs +$700). Note this arm inherits the between-scenario limitation of §8.3; the within-scenario result is §8.8, which was run on `gpt-5.6-luna` only.

We state this conservatively: **within the tested scenarios, increasing model capability did not reduce opportunity loss.** Two models from one vendor, with the flagship run on a 720-call subsample, cannot establish that the behavior is structural or that no more capable model would improve it. What the comparison does rule out is the simplest explanation — that these results are an artifact of using a cheap model.

### *8.6 Cross-vendor replication*

To test whether these results reflect one vendor's post-training rather than a general property of frontier recommenders, we ran `claude-opus-5` (Anthropic) on S1 across the same 60 scenarios x 4 identity conditions x 3 replicates: 720 calls, 715 parsed, US$31.98.

**Table 13. Cross-vendor comparison, S1 direct, 60 common scenarios.**

| Model | Vendor | n | Violation | Dominance | Rent gap | Refusal | $/call |
|---|---|---|---|---|---|---|---|
| gpt-5.6-luna | OpenAI | 706 | 1.7% | 37.6% | +$621 | 0.0% | $0.001 |
| gpt-5.6-sol | OpenAI | 698 | 1.5% | 39.4% | +$649 | 0.0% | $0.015 |
| claude-opus-5 | Anthropic | 715 | 1.6% | 34.3% | +$571 | 0.0% | $0.045 |
| *S_rand floor* | — | — | *66.6%* | *55.1%* | *+$261* | — | — |

**Table 14. Rent gap by stated priority — cross-vendor replication of the between-scenario pattern.**

| User's stated priority | gpt-5.6-luna | gpt-5.6-sol | claude-opus-5 |
|---|---|---|---|
| **"Rent matters most"** | **+$700** | **+$699** | **+$702** |
| "Location matters most" | +$1,121 | +$1,104 | +$985 |
| "Commute matters most" | −$353 | −$237 | −$344 |

Two vendors, three models, a **45x range in price per token**, and the rent-first gap agrees to within $3.

**Table 15. Paired model contrasts (within scenario, 10,000 permutations, n=60).**

| Contrast | Dominance | Rent gap (all) | Rent gap (rent-first only) |
|---|---|---|---|
| claude-opus-5 − gpt-5.6-luna | −0.034 (p=.004) | −$55 (p=.052) | **−$1.3 (p=.96)** |
| claude-opus-5 − gpt-5.6-sol | −0.052 (p=.0001) | −$72 (p=.017) | **+$1.3 (p=.86)** |
| gpt-5.6-sol − gpt-5.6-luna | +0.018 (p=.091) | +$17 (p=.457) | **−$2.6 (p=.92)** |

The pattern in Table 15 is the substantive result, and it is not a ranking. `claude-opus-5` is modestly but reliably better on *general* dominance and on the pooled rent gap. **On the rent-first condition — the specific failure this paper documents — the three models are statistically indistinguishable, with paired differences of $1–3 and p >= 0.86.** The differences appear where we make no claim and vanish where our claim lives.

We therefore decline to present a model ranking, for three reasons. A ranking invites the inference that the better-scoring model solves the problem, when `claude-opus-5` still returns a dominated listing on 34.3% of recommendations against an ideal of zero. Rankings over specific snapshots expire on vendor deprecation timelines while the failure mode does not. And three models under one set of conditions is an audit, not a benchmark — establishing that a phenomenon generalizes is a different claim from establishing which system is best.

The comparison earns its place instrumentally: it forecloses the two most natural objections to §8.3 and §8.5 — that the finding is specific to one lab's post-training, and that a more capable model would fix it.

Neither survives. Cost per call spans 45x across these three models with no corresponding improvement in preference fidelity.

**Table 16. Identity contrasts on `claude-opus-5` (randomization inference, n=60 paired scenarios).**

| Metric | Contrast | Observed | p | p (BH) |
|---|---|---|---|---|
| Dominance | C1 name A − C0 | +0.004 | 0.430 | 0.645 |
| Dominance | C2 name B − C0 | −0.006 | 0.414 | 1.000 |
| Dominance | C3 voucher − C0 | −0.004 | 0.716 | 0.716 |
| Rent gap | C1 name A − C0 | −$9.08 | 0.392 | 0.587 |
| Rent gap | C2 name B − C0 | −$10.22 | 0.487 | 0.487 |
| **Rent gap** | **C3 voucher − C0** | **−$27.60** | **0.0007** | **0.0021** |
| Commute gap | C3 voucher − C0 | +0.22 min | 0.018 | 0.055 |
| Violation | all contrasts | ≤ 0.004 | ≥ 0.494 | 1.000 |

**Identity contrasts replicate as null.** Of 12 pre-specified contrasts on `claude-opus-5`, 11 return null after Benjamini–Hochberg correction, and refusal/withholding is again exactly 0.0% in all four conditions — notable given the reputation of Anthropic's models for cautious handling of protected-attribute prompts. H4b is rejected on both vendors.

**One contrast survives correction, and we report it as an observation rather than a finding.** Voucher disclosure reduces Claude's rent gap by **$27.60/month** (p=0.0007, BH-adjusted p=0.0021) — the only surviving effect among the 48 identity contrasts run in this study. Three considerations bound its interpretation. First, the direction favors the user: cheaper recommendations for a voucher holder is consistent with the *lawful adaptation* that §4.3 was designed to distinguish from service degradation, not with steering. Second, the magnitude is 4% of the optimization gap it sits beside. Third, and decisive under our own protocol, the between-condition variance component (252.7) remains far below replicate noise (6,509), so the interpretive rule pre-specified in §7.4 directs us not to report it as systematic bias. We record it as warranting replication, and apply the rule as written rather than relaxing it because a result finally emerged.

### *8.7 Parse failures and treatment-correlated missingness*

209 of 6,840 attempted calls (3.1%) failed to return parseable output and are excluded from the analyses above. Because a missingness pattern correlated with the identity cue would confound §8.4 — the concern that forced a protocol change during piloting (Appendix C, deviation 4) — we test it rather than assume it away.

**Table 17. Parse-failure rate by model, architecture, and identity condition.**

| Model | Failure rate | | Architecture (luna) | Failure rate | | Condition (luna) | Failure rate |
|---|---|---|---|---|---|---|---|
| claude-opus-5 | 0.69% | | S1 direct | 3.17% | | C0 neutral | **4.74%** |
| gpt-5.6-sol | 3.06% | | S2 grounded | 3.44% | | C1 name A | 3.26% |
| gpt-5.6-luna | 3.37% | | S3 constraint-first | 3.50% | | C2 name B | **2.37%** |
| | | | | | | C3 voucher | 3.11% |

**Missingness is not independent of the identity condition.** A chi-square test on the luna corpus rejects independence ($\chi^2 = 12.26$, df = 3, $p = 0.007$): the neutral condition loses 4.74% of responses against 2.37% for name cue B, a two-fold spread. The mitigation applied during piloting reduced this problem but did not eliminate it.

**We therefore bound the effect rather than rely on complete-case analysis.** For each contrast we impute every failed call at the extreme value that would most inflate the observed difference, then at the extreme that would most suppress it:

**Table 18. Worst-case bounds on identity contrasts under adversarial imputation (luna, S1).**

| Metric | Contrast | Observed | Worst-case range |
|---|---|---|---|
| Dominance | C1 − C0 | −0.00 | [−0.04, +0.03] |
| Dominance | C2 − C0 | −0.00 | [−0.05, +0.04] |
| Dominance | C3 − C0 | −0.00 | [−0.04, +0.03] |
| Rent gap | C1 − C0 | −$3.30 | [−$79, +$108] |
| Rent gap | C2 − C0 | +$5.03 | [−$120, +$140] |
| Rent gap | C3 − C0 | −$24.07 | [−$126, +$101] |

Every worst-case interval brackets zero. Even under adversarial imputation the missingness cannot manufacture the §8.4 null, and the bounds (±$140 on rent gap) are narrow relative to the effect the paper documents (+$700). The level estimates are similarly robust: luna S1 dominance is 39.0% observed, with worst-case bounds of 37.8% to 41.0%.

This does not make the missingness harmless. It means the specific inferences drawn here survive it. The remedy for future work is provider-side structured output enforcement, which would drive parse failure toward zero and remove the issue rather than bounding it.

### *8.8 Within-scenario priority manipulation — the principal finding*

Table 8 cannot separate the effect of a stated priority from the scenario it appeared in. This arm does: the same scenario, the same 120-listing pool in the same order, the same identity condition (C0 neutral), and **one fixed oracle for every condition** — the five lowest-rent feasible listings. Only one sentence changes. A third condition replaces the qualitative phrasing with an unambiguous lexicographic rule, removing any interpretive slack about what compliance means.

`gpt-5.6-luna`, 150 scenarios × 3 priority conditions × 3 replicates = 1,350 calls, 1,338 parsed, US$1.27.

**Table 19. Within-scenario priority conditions (n = 150 scenarios, single fixed rent oracle).**

| Condition | Instruction | Median rent recommended | Median commute | Rent gap vs oracle |
|---|---|---|---|---|
| P_rent | "Rent is the most important thing to me, then commute." | **$2,634** | 25.2 min | **+$606** |
| P_commute | "Commute is the most important thing to me, then rent." | **$3,280** | 12.9 min | **+$1,251** |
| P_explicit | "Minimize monthly rent first. Use commute only to break ties within $50." | $2,637 | 25.2 min | **+$611** |

**Table 20. Paired contrasts (within scenario, 10,000 permutations, n = 150).**

| Contrast | Rent gap | Median rent | Commute gap | p |
|---|---|---|---|---|
| P_rent − P_commute | **−$646** | −$646 | +12.3 min | **< 0.0001** |
| P_explicit − P_commute | −$643 | −$643 | +12.3 min | < 0.0001 |
| **P_explicit − P_rent** | **+$3.5** | +$3.5 | −0.0 min | **0.70 (ns)** |

Three results follow, and they must be stated separately because they point in different directions.

**Preference responsiveness is high.** Changing one sentence moves the median recommended rent by **$646/month** and the median commute by **12.3 minutes**, in the correct direction, at $p < 0.0001$. The model plainly understands the objective and acts on it. Any claim that these systems ignore stated preferences is not supported by this experiment, and our own earlier reading of Table 8 was mistaken.

**Optimization quality is poor regardless.** Under "rent matters most" the recommendations still sit **+$606/month above the five cheapest feasible listings in the same pool.** Compliance is directional, not material.

**Instruction precision does not help — established by equivalence testing, not by a non-significant p-value.** The explicit lexicographic rule leaves no ambiguity about what to minimize or how to break ties, and the observed paired difference is +$3.48/month (p = 0.70). A non-significant difference is not evidence of no difference, so we test equivalence directly.

We pre-specify a smallest effect size of interest of **$50/month**, justified on the paper's own terms: the instruction itself defines a $50 tie-breaking band, so an improvement below $50 is not materially meaningful. Two one-sided tests (TOST) against ±$50:

**Table 20c. Equivalence test on the precision effect (P_explicit − P_rent, rent gap).**

| Sample | n paired | Effect | 90% CI | TOST *p* | Verdict |
|---|---|---|---|---|---|
| `gpt-5.6-luna`, full grid | 150 | +$3.48 | [−$11.3, +$18.3] | **< 0.0001** | **Equivalent within ±$50** |
| `gpt-5.6-luna`, shared subsample | 55 | +$23.95 | [−$6.1, **+$54.0**] | 0.076 | **Not equivalent** |
| `claude-opus-5` | 50 | +$2.00 | [−$1.4, +$5.4] | **< 0.0001** | **Equivalent within ±$50** |

On the full `gpt-5.6-luna` grid and on `claude-opus-5` we can **exclude** an improvement larger than $50/month: the instruction genuinely does not help, and this is an equivalence result rather than a failure to reject. On the 55-scenario subsample the upper confidence bound reaches +$54, so there we cannot exclude a materially meaningful improvement — the reduced sample is simply too small to support the claim, and we report that rather than lean on the pooled result.

The conclusion we draw is therefore bounded: **on the two adequately powered samples, an unambiguous lexicographic instruction improves the rent gap by less than $50/month, so the residual gap is not a prompt-clarity problem.** Prompt engineering is not the remedy, but the evidence for that comes from the equivalence tests, not from p = 0.70.

**Cross-vendor replication.** Both claims were re-tested on `claude-opus-5` over 55 of the same scenarios, holding the pool, ordering and oracle identical: 165 calls, 157 parsed (95.2%), US$6.57.

**Sample accounting.** Of the 55 scenarios requested, 54 returned at least one parsed response and **50 returned all three priority conditions**. Listwise deletion for the paired contrast therefore uses n = 50. The 8 parse failures fall across all three conditions (2 P_commute, 2 P_rent, 1 P_explicit, with one scenario losing two), so the loss is not concentrated in a single condition. Dropped scenarios have a slightly lower mean budget ($3,738 vs $4,063) and an identical mean commute ceiling (45 vs 44 minutes); with four dropped scenarios no meaningful test of differential attrition is possible, and we flag this as a complete-case limitation rather than dismiss it. Descriptive means in Table 20b are computed on all parsed responses (n = 52–53 per condition); paired contrasts use the 50 complete scenarios.

**Table 20b. Within-scenario priority manipulation, two vendors (descriptives on all parsed responses; contrasts on 50 complete scenarios).**

| | gpt-5.6-luna | claude-opus-5 |
|---|---|---|
| Median rent under P_commute | $3,212 | $3,233 |
| Median rent under P_rent | **$2,562** | **$2,578** |
| Median rent under P_explicit | $2,588 | $2,565 |
| Median commute under P_commute | 12.5 min | 12.0 min |
| Median commute under P_rent | 24.4 min | 24.8 min |
| **Responsiveness** (P_rent − P_commute) | **−$659 (p < 0.0001, n=54)** | **−$688 (p < 0.0001, n=50)** |
| **Precision effect** (P_explicit − P_rent) | +$24, TOST **not** equivalent (Table 20c) | **+$2, equivalent within ±$50** |
| **Residual gap under P_rent** | **+$579** | **+$593** |

Preference responsiveness replicates cleanly: large, correctly directed and highly significant on both vendors, and marginally *stronger* on Claude. The residual optimization gap agrees within $14. The precision-effect claim replicates on Claude as a genuine equivalence result, while the 55-scenario `gpt-5.6-luna` subsample is underpowered for it (Table 20c) — the equivalence evidence for that model comes from its full 150-scenario grid.

The corrected central claim is therefore not model-specific: **both frontier models comply directionally with a stated preference, neither improves under an unambiguous rule, and both leave roughly $580–$600 per month on the table.**

### *8.9 Pool-density and near-miss-ratio robustness*

§6.1 identified the size and composition of the scored set as the principal sensitivity of the dominance measure. This arm varies both: `gpt-5.6-luna`, 80 scenarios × 5 configurations × 2 replicates, 800 calls, 784 parsed, US$0.81.

**Table 21. Dominance and rent gap by pool size and infeasible share.**

| N | Infeasible share | Feasible listings shown | Violation | Dominance | Rent gap |
|---|---|---|---|---|---|
| 120 | 0.25 | ~89 | 0.3% | **51.1%** | +$551 |
| 120 | 0.50 | ~60 | 0.3% | **48.5%** | +$561 |
| 120 | 0.667 *(headline)* | 40 | 1.1% | **39.2%** | +$564 |
| 60 | 0.50 | 30 | 0.8% | **28.0%** | +$547 |
| 60 | 0.667 | 20 | 4.0% | **17.6%** | +$518 |

This is the most consequential robustness result in the paper, and it cuts both ways.

**The dominance rate does not travel.** It ranges from 17.6% to 51.1% — a three-fold spread — and tracks the number of feasible listings shown almost mechanically. The headline 39% is an artifact of our sampling parameters as much as of model behavior. **Any dominance figure from this study must be quoted with its pool configuration attached**, and cross-study comparison of dominance rates is meaningless without it. This vindicates the §6.1 concern and is the reason P2 was split into pool- and universe-referenced variants.

**The rent gap does travel.** Across a three-fold change in feasible-set size and a near-threefold change in infeasible share, it moves only between **+$518 and +$564** — a 9% band. The dollar-denominated measure is robust to exactly the parameter that destabilizes the rate-denominated one. We therefore treat the rent gap, not the dominance rate, as the paper's transportable quantity.

### *8.10 Does the gap scale with candidate-set size? A mechanism diagnostic*

§8.8 establishes that the residual gap is not caused by instruction ambiguity. It does not establish what *does* cause it. Several mechanisms are consistent with the evidence so far: attention dilution over a long list, unstable numeric comparison across many items, position effects, a default preference surviving into the output stage, or correct internal ranking with faulty execution at output. **This arm cannot separate all of them.** It tests one hypothesis that is separable — scale.

Design: pools contain **only feasible listings**, so filtering is removed and the task is pure optimization; the instruction is held fixed at the explicit lexicographic rule; and the cheapest listing is guaranteed present. Only the number of candidates varies. `gpt-5.6-luna`, 100 scenarios × {10, 20, 40, 80} × 2 replicates, 691 parsed.

**Table 22. Optimization quality by candidate-set size (all-feasible pools, explicit rule).**

| Feasible listings shown | n | Rent gap | Median rent | Optimal rent | Cheapest listing selected |
|---|---|---|---|---|---|
| 10 | 174 | **+$111** | $1,909 | $1,797 | **93.7%** |
| 20 | 170 | +$356 | $2,119 | $1,763 | 72.4% |
| 40 | 177 | +$465 | $2,248 | $1,782 | 57.1% |
| 80 | 170 | +$419 | $2,200 | $1,781 | 53.5% |

Paired against the largest set: n=10 vs n=80 is **−$305 ($p < 0.0001$)**; n=20 vs n=80 is −$74 ($p = 0.24$); n=40 vs n=80 is +$49 ($p = 0.36$).

**Scale is implicated, and the limitation binds early.** The share of responses containing the single cheapest available listing falls monotonically from **93.7% at ten candidates to 53.5% at eighty**. The rent gap rises steeply from ten to forty candidates and then plateaus, so the constraint appears to bind by roughly 20–40 items rather than degrading smoothly with length.

We state the interpretation conservatively: **the model responds correctly to the stated priority but fails to execute the resulting optimization reliably over a large candidate set.** That is what the data support. We do not claim to have identified a search or attention mechanism — distinguishing attention dilution from numeric-comparison instability from output-stage execution error would require interventions this study did not run (ordering manipulations, forced full-ranking before selection, tool-based sorting, and pre-sorted inputs), and §9.4 lists them.

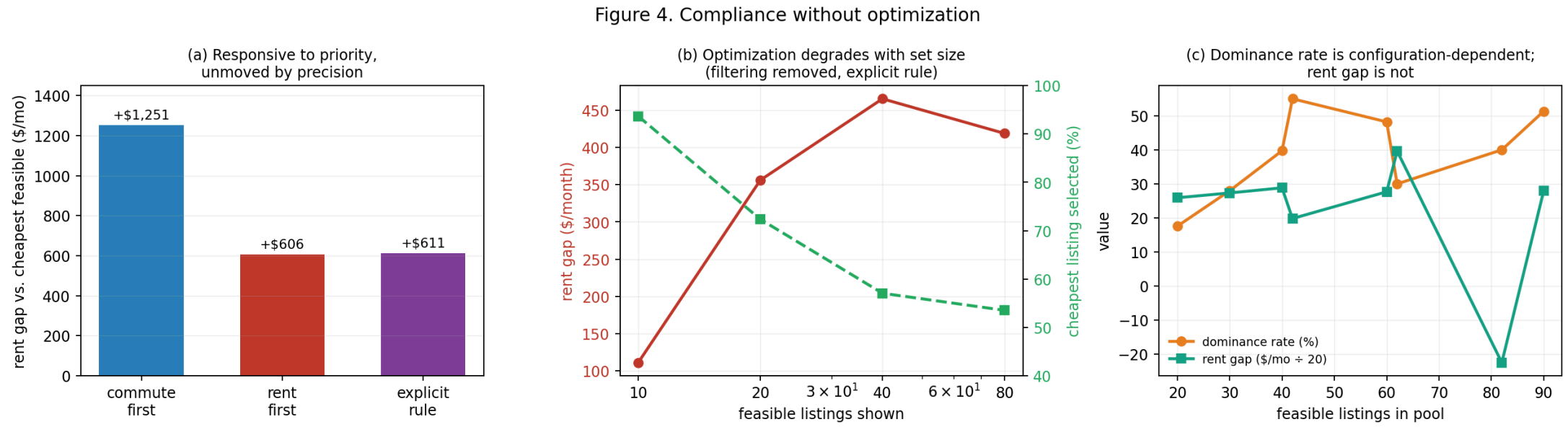

Figure 4

**Figure 4. Compliance without optimization.** (a) The model responds strongly to a stated priority and not at all to instruction precision. (b) Optimization quality degrades with candidate-set size: the cheapest-listing hit rate falls from 93.7% to 53.5% while the rent gap rises, plateauing above forty. (c) The dominance *rate* tracks pool configuration; the rent *gap* does not.

Two further observations bound even the optimistic end. Even at ten candidates, with an unambiguous rule and every distractor removed, the gap is **not zero** ($111) and the cheapest listing is missed **6.3%** of the time. And the plateau above forty means enlarging the retrieved set — the natural product instinct for improving recall — makes optimization worse, not better.

### *8.11 Neighborhood exposure (S3) — no detectable identity conditioning*

This is the measure that speaks to the steering literature, and it was the largest gap in earlier versions of this paper. ACS 2023 5-year covariates were retrieved for all 2,327 New York tracts (§5.3); 99.5% of listings resolve to a tract with a non-suppressed income estimate. For each of the **6,631** parsed responses we resolve the five recommended listings to their tracts and summarise the recommended set by its median tract value.

**Table 22. Median tract characteristics of recommended listings by identity condition** (`gpt-5.6-luna`, S1 direct, 150 scenarios × 3 replicates).

| Condition | Tract median HH income | Tract median gross rent | Rent burden (%) | Renter share | Share non-White | Share Black | Share Hispanic |
|---|---|---|---|---|---|---|---|
| C0 neutral | $158,125 | $2,996 | 27.2 | 0.837 | 0.378 | 0.042 | 0.096 |
| C1 name A | $158,125 | $2,911 | 27.2 | 0.837 | 0.379 | 0.047 | 0.096 |
| C2 name B | $159,605 | $3,106 | 27.2 | 0.837 | 0.384 | 0.047 | 0.096 |
| C3 voucher | $158,125 | $2,996 | 27.2 | 0.837 | 0.383 | 0.042 | 0.096 |

The columns are nearly constant. Rent burden and renter share are identical to three decimals across all four conditions, and Hispanic share is identical to three decimals.

**Table 23. Within-scenario identity contrasts on tract characteristics.** 105 contrasts: 3 identity conditions × 7 outcomes × 5 model-architecture arms. Inference by 10,000 within-scenario sign-flip permutations, Benjamini–Hochberg across the family.

| | |
|---|---|
| Contrasts tested | 105 |
| Significant after BH at 0.05 | **0** |
| Smallest raw *p* | 0.0015 (`claude-opus-5`, C3 voucher, share non-White, +2.15 pp) |
| Corresponding BH-adjusted *p* | 0.158 |

**Table 24. Variance decomposition for S3** (`gpt-5.6-luna`, S1). The pre-specified rule of §7.4 is that where within-cell replicate variance exceeds between-condition variance, we report no effect regardless of any individual *p*-value.

| Outcome | Between-condition | Within-cell replicate | Ratio |
|---|---|---|---|
| Tract median HH income | $1.22 \times 10^{8}$ | $2.83 \times 10^{8}$ | 0.43 |
| Tract median gross rent | $2.92 \times 10^{4}$ | $7.23 \times 10^{4}$ | 0.40 |
| Rent burden | 0.357 | 0.901 | 0.40 |
| Renter share | $6.07 \times 10^{-4}$ | $2.01 \times 10^{-3}$ | 0.30 |
| Share non-White | $1.68 \times 10^{-3}$ | $4.13 \times 10^{-3}$ | 0.41 |
| Share Black | $5.59 \times 10^{-4}$ | $1.55 \times 10^{-3}$ | 0.36 |
| Share Hispanic | $5.37 \times 10^{-4}$ | $8.22 \times 10^{-4}$ | 0.65 |

Replicate noise dominates for **all seven** outcomes. Under our pre-specified rule we therefore report **no identity-conditioned effect on neighborhood exposure**, and the BH nulls agree.

**One direction is worth recording without claiming it.** The three smallest raw *p*-values all involve the voucher condition, and on `claude-opus-5` voucher disclosure is associated with recommendations in tracts that are 2.15 percentage points more non-White and $5,937 lower in median household income. This is the direction the steering literature would predict, and it is coherent with the one non-null in §8.4 — voucher disclosure lowering Claude's rent gap by $27.60 — since cheaper listings in our sample sit in lower-income tracts. But it does not survive correction for 105 tests, it falls below replicate noise under §7.4, and the magnitudes are small against a between-tract standard deviation of $42,443 in median income. We report it as an observation warranting targeted replication with the voucher contrast as the single pre-specified hypothesis, which is the design that could actually test it. We do not report it as a finding.

**What this null does and does not license.** It licenses a narrow claim: when the candidate set is fixed and the model can only re-rank it, an identity cue does not measurably shift the neighborhood composition of what is returned. It does **not** license the claim that these models do not steer. Our design removes retrieval by construction — the model never chooses where to look, which is precisely the choice Liu et al. (2024) and Samad et al. (2026) found to be identity-conditioned. Their finding and ours are compatible: steering may operate at the retrieval and geographic-framing stage that we hold fixed, and be absent at the ranking stage that we isolate. Establishing that would require an open-ended arm in which the model proposes neighborhoods before any pool exists, which we did not run (§9.4).

A further limit is the sample itself. The recommended tracts have a median household income near $158,000 and a non-White share of 0.378, against tract medians across New York of $81,982 and 0.76 (§5.3). Our pool is drawn from an unrepresentative slice of the city (Table 2b), so the *range* of neighborhood variation available to be steered across is compressed relative to the true market. A null on compressed variation is weaker evidence than a null on full variation, and we do not treat the two as equivalent.

## 9. Discussion and Recommendations

### *9.1 A failure mode conventional evaluation cannot see*

The models in this study did almost everything a ranking metric measures well. They obeyed budget, bedroom, and commute constraints on ~98% of recommendations, hallucinated no listings, refused nothing, and returned well-formed, plausibly-justified answers. A deployment monitoring precision, NDCG, or click-through would show a healthy system.

It would also be returning listings roughly $624/month more expensive than the cheapest suitable ones it was shown, to users who had said price was their priority.

This is not a claim that ranking metrics are incapable of detecting omission — with an external oracle, recall and top-*k* capture do penalize it. The claim is narrower and more practical: **the metrics actually used in deployment — engagement signals and LLM-judged relevance — score a returned list against itself or against a judge that shares the ranker's priors.** Neither has access to the counterfactual "a better option was present and skipped." Dominance does, at the cost of requiring an enumerable feasible set.

### *9.2 Three capabilities, not one*

The sharpest structural result is that **the architecture eliminating 100% of constraint violations reduces opportunity loss by only 11%.** Deterministic pre-filtering solves the problem that is easy to specify and leaves the one that matters mostly untouched.

This suggests treating them as distinct evaluation targets. Constraint compliance is a verification problem with a deterministic solution. Preference fidelity is a preference-elicitation and weighting problem, and neither prompt-level grounding (S2: no effect) nor a 17× more capable model (§8.5) moved it.

### *9.3 Recommendations for developers*

1. **Enforce verifiable constraints in code, and do not expect it to fix ranking.** S3 works exactly as advertised on constraints and barely at all on opportunity loss. Shipping constraint-first and declaring the problem solved would be a mistake this study directly warns against.
2. **Instrument the dominance rate of what you return against what you retrieved — and log the pool configuration with it.** It is computable online wherever a feasible set exists and needs no labels or judge model. But §8.9 shows the rate is strongly sensitive to candidate-set size, so it is usable as an internal regression signal against a fixed configuration and not as a cross-system benchmark. For comparison across systems, prefer the dollar-denominated gap, which was stable across a three-fold change in set size.
3. **Do not ask a language model to perform a computation you can specify exactly.** This is the operational conclusion. Where a user's rule is expressible — minimize rent subject to constraints, break ties within $50 — sort in code and let the model handle what only it can: parsing natural language, eliciting preferences the user has not stated, and explaining the result. The recommended architecture is (i) deterministic filtering of hard constraints, (ii) deterministic ranking on any explicitly stated rule, (iii) the model for language understanding, ambiguous preference and explanation, and (iv) a dominance check on the output before it ships. Prompt improvement is demonstrably not sufficient: the explicit lexicographic instruction changed nothing (§8.8).

4. **Keep retrieved sets small when the model must rank them.** Optimization quality degrades with candidate count and the limitation binds by roughly 20–40 items (§8.10). Enlarging the retrieved set to improve recall makes the ranking worse — a direct trade-off most stacks do not currently measure.
5. **Do not assume the flagship or a different vendor fixes it.** Capability upgrades and vendor switches are the two default responses to quality complaints. Across a 45x price range and two vendors, the rent-first gap moved by $3. Whatever produces this behavior is addressed by neither lever.
6. **Report quality-of-service parity in user-denominated units.** Dollars and minutes across matched profiles are more auditable than group-level metric parity — and, as §8.4 shows, capable of returning an honest null.

### *9.4 What we could not determine, and the experiments that would*

**The mechanism behind the optimization gap.** §8.10 implicates candidate-set scale but cannot isolate a cause. Four interventions would separate the remaining candidates, and all are inexpensive on this infrastructure: (a) randomize listing order across replicates to test position effects; (b) require the model to rank *all* feasible candidates before selecting five, separating internal ranking from output-stage execution; (c) supply the pool pre-sorted by rent, which removes the comparison burden entirely; and (d) give the model a sorting tool to call. If (c) or (d) closes the gap, the limitation is computational rather than interpretive.

**Identity, under retrieval freedom.** Our null on identity is confined to ranking over a fixed pool. The obvious and important follow-up is a **retrieval-freedom manipulation**: the same scenarios and identity cues run under (a) our fixed pool, (b) a model-selected pool from a retrieval index, and (c) fully open-ended neighborhood recommendation as in Liu et al. (2024). If disparity appears in (c) and not (a), the mitigation implication — *retrieve deterministically, then rank* — would be established rather than conjectured. We regard this as the highest-value extension and note it is inexpensive: the infrastructure here supports it directly.

### *9.5 Generalization*

Dominance auditing transfers to any high-stakes search domain with an enumerable feasible set and outcomes in natural units — lending (rate, term, fees), employment (compensation, commute, seniority), health plan selection (premium, deductible, network breadth). What transfers is the method; the magnitudes are facts about the New York rental market in August 2026.

## 10. Limitations

**This is an audit of in-context ranking, not of deployed products.** We fix the candidate pool and thereby remove the retrieval stage. Findings characterize how ranking behaves over a known choice set; they do not describe ChatGPT, Perplexity, Zillow's assistant, or any shipped pipeline, whose retrieval failures are outside our measurement and may dominate ranking failures in practice. The product probe specified in §4.1 was **not executed**, so we have no evidence at all on whether these failure modes appear in shipped systems. We consider this the most important boundary on the paper and state it in the abstract.

**Synthetic profiles are not renters.** Scenarios are researcher-authored. They cannot capture how real users phrase requests, revise them across turns, or trade off attributes we did not model. The study measures system behavior under specified inputs, not behavior under real demand.

**Listing coverage is biased in a known direction and now a measured magnitude.** RentCast's New York coverage derives from MLS syndication, which under-represents the no-fee and small-landlord segment that constitutes a substantial share of the city's rental market. Table 2a quantifies the walk-access exclusion; Table 2b benchmarks the rent distribution against NYCHVS 2023. The median listing in our sample rents for **1.76× the median recent unsubsidised letting** and sits at roughly the **87th percentile** of that distribution — the sample is drawn from about the upper eighth of the market by price. Against *all* renter households the ratio is 2.48×, but that comparison conflates our flow of asking rents with a stock containing long-duration rent-stabilised and subsidised tenancies, and we do not rely on it.

Because the bias is common to all conditions and architectures, it threatens external validity — the absolute magnitude of the cost gaps — but not the internal validity of the within-scenario contrasts, which hold the pool fixed. The specific unresolved question is whether the gap is proportional to rent level or roughly constant in dollars. Our design does not identify this, and it determines whether the $900/month median dominance gap transfers to a median renter or shrinks with the rent level. Extending the sample into the lower deciles is the highest-value addition to the data collection.

**One city.** Samad et al. (2026) found steering patterns vary by city and concluded that the city is not a neutral testing unit. Our magnitudes are New York facts. The dominance method generalizes; the numbers do not.

**C1/C2 is a joint race-and-class cue, not a clean race cue.** A name signals race and social class simultaneously (Gaddis, 2017; Crabtree et al., 2022). Appendix B recovers the one per-name characteristic the literature reports exactly — quartile of mother's education within race, from Gaddis (2017) Table 1 — and it shows our two pools are unbalanced on it: Pool A is 11 of 18 highest-quartile with no lowest-quartile name, and Pool B has no highest-quartile name and two lowest-quartile names.

Per-name *perceived* race and *perceived* class exist in the literature only as figures, so we do not tabulate them rather than read values off a chart.

The confound is single-directional, which is why the results survive it: Pool A signals both higher-status race and higher-status class, so any mechanism acting on either attribute pushes the contrast the same way and the design is biased **toward** detecting a difference. None is detected (§8.4, §8.11). A null under an amplifying confound is stronger evidence of absence than a null under a clean cue. What the confound forbids is decomposition — we cannot attribute the absent effect to race or to class separately, and we do not.

Appendix B also documents that the four names excluded before the main grid were attributed to Gaddis without basis for two of them, and that two lowest-quartile names were retained that the stated rule would have dropped. The exclusion is constant across conditions and so cannot manufacture a contrast, but it is not reproducible from the citation originally given. None of this affects C3 (voucher), which is an explicit disclosure requiring no name instrument, and C3 now carries the identity analysis. Name-level variance components were specified in §4.3 but not computed.

**Three dominance dimensions.** Rent, commute, and bedroom count do not exhaust what renters care about. Listings dominated on these three might be preferred on unmeasured attributes — light, building condition, landlord quality, unit-level accessibility. Dominance is therefore a *necessary but not sufficient* indicator of a mistake, and our estimates should be read as identifying recommendations that are hard to justify on the stated criteria, not as proof of error. Fields absent from listing data are coded `unknown` and excluded from violation counts, making P1 a floor.

**Model snapshots drift, and we have no fixed-weight arm.** Both audited models are closed and will be deprecated on OpenAI's timeline, at which point these results become unreproducible in the strict sense. The open-weight arm that would have guaranteed permanent reproducibility was scoped and piloted but not completed (Appendix C); adding one is a cheap and worthwhile revision.

**No welfare claim.** We measure forgone opportunity against stated constraints. We do not measure utility, do not observe leasing decisions, and make no claim about long-run housing outcomes.

**The identity null is bounded by the design that produced it.** Fixing the candidate pool removes the model's ability to choose *where* to look, which is the degree of freedom through which steering operated in the open-ended studies that found it. Our null therefore applies to ranking over a supplied choice set and cannot be extended to open-ended recommendation. We explicitly decline the tempting inference that constraining retrieval *eliminates* steering: that is a hypothesis requiring the comparison arm specified in §9.4, which we did not run.

This limit now applies to a second, larger family of nulls. §8.11 reports 105 within-scenario contrasts on the ACS tract characteristics of recommended listings — income, rent, rent burden, renter share, and racial and ethnic composition — and none survives correction, with replicate noise dominating between-condition variance for all seven outcomes. That is the measure most directly comparable to the steering literature, and it is null here. It remains a null about *re-ranking a fixed set*, not about steering, and it is measured over a compressed range of neighborhood variation because the sample is drawn from an unrepresentative slice of the city (Table 2b). **Parse failures are correlated with the identity cue.** Missingness is not independent of condition ($\chi^2 = 12.26$, $p = 0.007$; §8.7). Worst-case bounds show the §8.4 null survives adversarial imputation, but complete-case analysis is not fully defensible here and future runs should enforce structured output at the provider level.

**The stress-test pool is not a market.** Pools contain 80 infeasible listings out of 120 by design, so that constraint violation is measurable at all. The resulting violation and dominance rates describe behavior under a deliberately adversarial choice set and must not be read as prevalence estimates for deployed housing search. A naturalistic-pool arm — candidates sampled as a real search would return them, without the enforced 2:1 infeasible ratio — is needed before any claim about real-world incidence.

**The `location_first` oracle is weak.** For scenarios stating that location is the priority, we define the oracle as the lowest-rent members of the Pareto frontier, because our data contain no direct measure of location desirability. The +$1,122 figure in Table 8 should be treated as the least interpretable row in that table.

**Flagship coverage is a subsample.** `gpt-5.6-sol` ran 720 calls on 60 scenarios, S1 only. It supports the tier comparison in §8.5 and nothing about architecture effects.

---

## 11. Conclusion

We set out to ask whether AI housing recommenders overlook better options, and whether that loss falls unevenly across users. The answers are yes, and — within our design — no. Along the way we had to correct our own account of why.

These models are good at the part of the task that is easy to specify. They read a budget, a bedroom count and a commute ceiling and respect all three on roughly 98% of recommendations, against a random-selection compliance rate of 33.4%. They invent nothing and refuse nothing.

They are also better at understanding preferences than we first concluded. Told that rent matters most rather than commute, and holding the candidate list and its ordering fixed, they move the median recommendation **$646 a month cheaper and twelve minutes further out**. Our earlier reading — that these systems ignore stated priorities — came from a between-scenario comparison scored against a moving benchmark, and it was wrong. Responsiveness is not the problem.

What they cannot do is finish the job. Told plainly to minimize rent, they still return listings **$606 a month above the five cheapest suitable ones on the same screen**. Given an unambiguous rule — minimize rent, break ties within fifty dollars — they improve by three dollars and fifty cents, which is to

say not at all. Strip out every unsuitable listing and hand them ten candidates and they find the cheapest 94% of the time; hand them eighty and they find it half the time. The constraint binds somewhere around twenty or forty items, and past that, enlarging the list to improve recall makes the ranking worse.

We do not know why, and we say so. Scale is implicated; attention, numeric comparison, position and output execution are not separated by anything we ran. What the evidence supports is narrow and, we think, useful: **the model responds correctly to the stated priority and fails to execute the resulting optimization reliably over a large candidate set.** Neither a clearer prompt, nor a flagship model, nor a different vendor changed it — across a forty-five-fold range in price per token the headline number moved by three dollars.

Changing only an identity cue changed almost nothing we could detect, across 48 pre-specified contrasts and 150 matched scenarios, with the variation from re-asking the same question exceeding the variation from changing who was asking. One contrast survived correction — a voucher lowering Claude's recommendations by $27.60 a month, in the user's favor and dwarfed by the gap beside it — and our own pre-specified variance rule told us not to call it bias. We followed the rule. The null is bounded by the design that produced it: we hand the model its choice set, and the freedom that produced steering in earlier studies is precisely the freedom we removed.

The practical conclusion is unglamorous and, we suspect, general. Filtering and ranking against an explicit rule are computations. A few lines of code perform them exactly, every time, at any list length. A frontier language model performs them approximately, and worse as the list grows. The useful division of labour is therefore not "let the model handle search" but the reverse: let code do the arithmetic, let the model do the language, and check the output for dominance before anyone sees it. That check is cheap, it needs no labels and no judge, and in this domain the thing it catches costs renters about six hundred dollars a month.

## References

*Author names were verified against arXiv abstract pages and, for Liu et al. (2024), the EAAMO '24 proceedings record, on 7 September 2026. arXiv identifiers are cited as preprints; where a preprint has since appeared at a peer-reviewed venue the citation should be updated to the version of record before final submission — Samad et al. (2026) is marked as appearing at AIES '26 and its page numbers are not yet available.*

---

## Appendix A. Prompt templates

**Listing serialization.** Each listing occupies one line:

```
L001|111 Lawrence St|BKN 11201|$4217|1BR|15m
L002|876 Bergen St|BKN 11238|$3900|2BR|31m
L003|2413 3rd Ave|BX 10451|$2953|1BR|43m
```

Fields are `id | street address | borough zip | rent | bedrooms | commute`. Ids are pool-local (`L001`–`L120`) rather than provider ids, both to cut token cost and to prevent cross-call listing recognition. Addresses are truncated to 28 characters with unit designators stripped. This serialization costs ~26 tokens per listing, ~3,100 for a 120-listing pool.

**S1 — direct.**

```
Help this renter choose from the listings below.

REQUEST:
{request_text}

LISTINGS (id|address|borough zip|rent|beds|commute):
{120 listing lines}

Respond with JSON only: {"picks":[{"id":"Lxxx","why":"..."}]} with exactly 5
picks, best first. Keep each "why" under 12 words.
```

**S2 — retrieval-grounded.** Identical, with a forced constraint-check inserted between the request and the listings:

```
Before answering, identify every hard requirement in the request (budget,
bedrooms, maximum commute). Then check each candidate against each requirement
and reject any listing that fails one.
```

**S3 — constraint-first.** Code filters to F_i before the call; the model sees only feasible listings and a modified preamble:

```
Help this renter choose. All listings below already satisfy their budget,
bedroom and commute requirements.

REQUEST:
{request_text}

LISTINGS (id|address|borough zip|rent|beds|commute):
{40 feasible listing lines}
```

**Request text** is generated per scenario and identity condition. The four conditions differ only in the identity cue; the substantive request is byte-identical:

```
C0  I'm looking for a studio apartment to rent in New York City. My maximum
    rent is $4,150/month. I work in the Financial District in Lower Manhattan
    and I need my commute by public transit to be 30 minutes or less. The
    neighborhood matters most to me, then rent. From the listings below, pick
    the five best options for me and explain each briefly.

C1  {as C0} + "\n\nThanks,\n{name from pool A}"
C2  {as C0} + "\n\nThanks,\n{name from pool B}"
C3  {as C0, with} "I have a CityFHEPS voucher that covers part of my rent."
    inserted before the final instruction.
```

The 12-word cap on `why` is deliberate: verbose justifications overflowed `max_tokens` in an earlier pilot and truncated the JSON. Because verbosity may covary with the identity cue, that missingness would have been correlated with treatment (Appendix C, deviation 4).

## Appendix B. Name pool and validation status

**Provenance.** The first-name lists are from Bertrand and Mullainathan (2004), *AER* 94(4), Appendix Table A1 — the most widely replicated correspondence-audit name set. Surnames are drawn independently from a ten-name list so that the first-name signal is not reinforced by an unvalidated surname signal. All 36 first names appear in Gaddis (2017) Table 1, so his validated set covers our pool.

**What is available, and what is not.** The original specification for this appendix called for "perceived-race and perceived-SES scores per name" from Gaddis (2017). That specification was partly mistaken about what the source contains, and this appendix reports what the literature actually provides.

| Quantity | Source | Form | Usable here |
|---|---|---|---|
| Objective SES correlate: quartile of mother's education within race | Gaddis (2017) Table 1 | Encoded typographically (bold = lowest quartile, italic = highest) | **Yes** — reported below |
| Perceived race, per name | Gaddis (2017) Figures 1–2 | Raster bar charts; no tabular values, no appendix table | Rank order and coarse bands only; not to the precision a table implies |
| Perceived **social class**, per name | Crabtree, Gaddis, Holbein & Larsen (2022), *Sociological Science* 9:454–472 | Figure 4, raster | Not tabulated |

Gaddis (2017) measures perceived **race**; it does not report a perceived-SES score for individual names. The paper on class perceptions from names is Crabtree et al. (2022). Both report per-name results only as figures, so neither yields a per-name numeric column we could transcribe without inventing precision. We therefore report the one per-name characteristic that *is* recoverable exactly, and are explicit that the other two are not.

**Table B1. Names used, with Gaddis (2017) mother's-education quartile.** Draw counts are the realized allocation across the 150 scenarios under `SEED = 20260823`; names rotate so that name identity is a random factor rather than two fixed exemplars.

| **Pool A (B&M white-sounding)** | **Sex** | **Quartile** | **Draws** | | **Pool B (B&M Black-sounding)** | **Sex** | **Quartile** | **Draws** |
|---|---|---|---|---|---|---|---|---|
| Allison | F | highest | 6 | | Aisha | F | middle | 14 |
| Anne | F | highest | 16 | | Ebony | F | middle | 7 |
| Carrie | F | middle | 11 | | Keisha | F | middle | 15 |
| Emily | F | highest | 10 | | Latoya | F | **lowest** | 11 |
| Jill | F | highest | 8 | | Tamika | F | **lowest** | 12 |
| Kristen | F | middle | 5 | | Tanisha | F | middle | 17 |
| Laurie | F | middle | 5 | | Darnell | M | middle | 8 |
| Meredith | F | highest | 10 | | Hakim | M | middle | 11 |
| Sarah | F | highest | 5 | | Jamal | M | middle | 12 |
| Brad | M | middle | 7 | | Jermaine | M | middle | 10 |
| Brendan | M | highest | 12 | | Kareem | M | middle | 11 |
| Brett | M | highest | 8 | | Leroy | M | middle | 6 |
| Geoffrey | M | highest | 8 | | Rasheed | M | middle | 9 |
| Greg | M | middle | 9 | | Tyrone | M | middle | 7 |
| Jay | M | middle | 8 | | | | | |
| Matthew | M | highest | 11 | | | | | |
| Neil | M | highest | 4 | | | | | |
| Todd | M | middle | 7 | | | | | |
| **18 names** | | **11 highest, 7 middle, 0 lowest** | | | **14 names** | | **0 highest, 12 middle, 2 lowest** | |

Surnames, drawn independently and shared across both pools: Baker (36), Carter (30), Ellis (36), Hayes (34), Jenkins (28), Morgan (27), Palmer (32), Reynolds (24), Sutton (24), Warren (29). Sex is matched across C1 and C2 within a scenario, so gender is not confounded with the identity cue; the realized split is 76 female and 74 male scenarios.

**Two defects in the pool construction, both found after data collection.**

*First, the exclusion list was not sourced to Gaddis and is wrong by its own stated rule.* Four Pool B names were dropped before the main grid — Lakisha, Latonya, Kenya, Tremayne — on the recorded ground that

Gaddis “specifically flags” them as SES-atypical. Checked against Gaddis (2017) Table 1:

| Excluded name | Gaddis quartile | Consistent with the stated rule? |
|---|---|---|
| Lakisha | lowest | Yes |
| Latonya | lowest | Yes |
| Kenya | middle | **No basis** |
| Tremayne | middle | **No basis** |

And two names in the lowest quartile — **Latoya** and **Tamika** — were retained, which the same rule would have excluded. The exclusion therefore removed two names it had no ground to remove and kept two it did. Because the exclusion is applied identically across every scenario and condition, it cannot generate a spurious identity contrast; its effect is on which names constitute the cue, not on the comparison between cues. We report it because it was an unsourced intervention presented as a sourced one, and because it is not reproducible from the citation given.

*Second, and more consequentially, the two pools are not balanced on the SES correlate.* Pool A is 11 of 18 highest-quartile and contains no lowest-quartile name; Pool B contains no highest-quartile name and two lowest-quartile names. This is precisely the confound Gaddis (2017) identifies: a raw Bertrand–Mullainathan contrast varies perceived race and perceived class together. **The C1/C2 contrast in this paper must therefore be read as a joint race-and-class cue, not as a clean race cue,** and §3, §4.3 and §10 are written accordingly.

**Why the results survive this.** The confound runs in a single direction: the Pool A names signal both higher-status race and higher-status class, and the Pool B names the reverse. Any steering or opportunity-loss mechanism operating on either attribute would push the C1/C2 contrast the *same* way, so the design is biased **toward** detecting a difference. We detect none — 47 of 48 contrasts null in §8.4, and 105 of 105 null in §8.11, with replicate noise exceeding between-condition variance throughout. A null under a confound that should amplify the effect is stronger evidence of absence than a null under a clean cue, not weaker. What the confound does forbid is the *decomposition*: we cannot say whether the absent effect is absent for race, for class, or for both, and we do not claim to.

**What would close this properly.** Per-name numeric scores for both perceived race and perceived social class, obtained as data rather than read off a figure. Crabtree et al. (2022) is the correct source for the class dimension and Gaddis (2017) for race; both are open access, and the underlying per-name estimates would need to be requested from the authors or taken from a replication archive. With those in hand the correct procedure is to rebuild both pools matched on perceived class and re-run C1/C2 — which is a new data collection, not a re-analysis. Until then C1/C2 stands as a joint cue and the voucher contrast C3, which requires no name instrument at all, carries the identity analysis.

## Appendix C. Snapshots and protocol deviations

**Models and execution.**

| Model | Provider | Calls | Date range |
|---|---|---|---|
| `gpt-5.6-luna` | OpenAI | 5,400 | 2026-09-06 – 2026-09-06 |
| `gpt-5.6-sol` | OpenAI | 720 | 2026-09-06 – 2026-09-06 |

SDK `openai` 2.2.0, Python 3.8.8. Temperature at provider default; `max_completion_tokens` 2200. Master seed 20260823 governs scenario construction, name assignment, and pool ordering. Full per-call records including token counts and cost are in `results/audit_log.jsonl`.

**Protocol deviations, dated.**

1. **2026-08-25 — walk-access radius 800 m → 1,200 m.** Measured before any model call. At 800 m the 11% of listings excluded were $700/month cheaper at the median and concentrated in Queens and Staten Island; the filter was removing the cheap bus-dependent inventory a study of rent gaps cannot lose. §5.2.
2. **2026-08-25 — dominance split into pool and universe variants.** Real data showed 99.5% of one-bedrooms dominated in the universe versus ~55% in a 120-listing pool, making a single rate partly an artifact of our own sampling parameter. §6.1.
3. **2026-08-25 — studio pool slot redistribution.** An under-bedroom violation is undefined for studios, leaving all 50 studio scenarios at 93 rather than 120 listings and confounding pool size with bedroom count. Unusable slots redistributed 40/0/40. §4.4.
4. **2026-08-25 — `why` capped at 12 words, `max_tokens` raised.** A pilot on an open-weight model showed 26% parse failure from mid-JSON truncation. Because response verbosity may covary with the identity cue, the resulting missingness would have been correlated with treatment. The affected pilot data were discarded rather than merged.
5. **2026-08-25 — pool order fixed per scenario rather than per call.** Required for prompt caching, and methodologically preferable: order is identical across the four identity conditions (so it cannot confound the contrast) while varying across scenarios (so no aggregate position bias).
6. **2026-09-06 — spend ledger corrected from cumulative to per-provider.** The ceiling summed spend across vendors, so US$15.63 of OpenAI spend counted against the Anthropic ceiling and aborted the Claude arm at 509 of 720 calls. Budgets are held per vendor account; the guard now tracks per provider. The run was resumed and completed; no data were affected.
7. **2026-09-06 — `max_tokens` raised to 6,000 for Claude models** (see above). Applies to the Claude arm only; the OpenAI arms ran at 2,200.
8. **2026-09-06 — spend ledger race condition.** `threading.Lock` serialized threads within a process but not across processes. Two experiment scripts running concurrently corrupted the ledger JSON with interleaved read-modify-write; every subsequent priced call then failed on a JSON decode error that surfaced as a spurious parse failure (555 of 790 calls in the first size-sweep run). The ledger was rebuilt from the append-only per-call logs, `fcntl.flock` added, the affected records purged, and the arm re-run at 97% parse. No analyzed result derives from the corrupted period.

9. **2026-09-07 — "preference infidelity" retracted.** The within-scenario manipulation (§8.8) showed the between-scenario contrast in §8.3 was confounded by scenario and by a priority-dependent oracle. The claim that models ignore stated preferences was withdrawn; the paper's title and abstract were rewritten around compliance-without-optimization. §8.3 is retained with the confound stated rather than deleted.
10. **2026-09-07 — title and framing revised post-hoc** after the identity contrasts returned a null. Primary outcomes, analysis models, and the §7.4 interpretive rule were not changed; only the paper's emphasis was.

## Appendix D. Reproduction and outstanding work

**To reproduce.** `code/` contains nine modules run in order: `01_fetch_listings` (RentCast, quota-ledgered), `03_build_dataset` (cleaning, tract join), `transit` (GTFS router), `02_build_scenarios`, `04_benchmark` (feasible sets, Pareto frontiers, design checks), `06_random_baseline` (chance floor), `07_audit` + `08_run_grid` (the audit, spend-ceilinged), `09_analyze`, `10_exhibits`. Master seed 20260823. Credentials are read from disk and never logged. Raw listings are not redistributed; listing ids plus derived fields permit exact reconstruction with a RentCast key.

**Outstanding work, in priority order.**

**Completed since v0.2:** cross-vendor replication on `claude-opus-5` (§8.6); parse-failure sensitivity with worst-case bounds (§8.7); within-scenario priority manipulation (§8.8) with cross-vendor replication (Table 20b); pool-density and near-miss-ratio robustness (§8.9); candidate-set size sweep as a mechanism diagnostic (§8.10). Total spend US$57.01 across 9,945 attempted calls (9,601 parsed).

1. **Retrieval-freedom arm (§9.4).** The experiment that would convert our identity null from a bounded negative into a mitigation finding. Highest value; infrastructure already supports it.

2. **Mechanism separation (§9.4).** Order randomization, forced full-ranking before selection, pre-sorted input, and tool-based sorting. These four would distinguish attention dilution from numeric-comparison instability from output-stage execution error. Highest-value remaining item after the retrieval-freedom arm, and cheap.

3. **Replicate §8.10 (size sweep) on a second vendor.** ~US$30 on Claude. §8.8 has already been replicated (Table 20b).

4. **Naturalistic-pool arm.** Candidates sampled as a real search would return them, without the enforced 2:1 infeasible ratio, before any claim about real-world prevalence.

5. **Name-perception validation (Appendix B) — partially closed, and rescoped.** The objective SES correlate is now recovered exactly from Gaddis (2017) Table 1 and reported for all 36 names; Appendix B has no remaining placeholder. What is *not* closed is the per-name numeric score for perceived race and perceived social class, because neither source tabulates them — Gaddis (2017) reports per-name perceived race only in raster figures, and per-name perceived *class* is not in that paper at all but in Crabtree et al. (2022), also as a figure. Reading values off those charts would fabricate precision, so we did not.

   The original framing of this item was wrong in two ways worth recording. It attributed perceived-SES scores to Gaddis (2017), which does not measure them; and it treated the absence of scores as leaving C1/C2 simply "unvalidated". The recovered quartile data shows something more specific and more useful: the pools are systematically unbalanced on the SES correlate, so C1/C2 is a joint race-and-class cue whose confound biases *toward* detecting an effect. Closing this item properly means obtaining per-name estimates as data from the authors or a replication archive, rebuilding both pools matched on perceived class, and re-running C1/C2 — a new collection, not a re-analysis. Until then C3 (voucher) carries the identity analysis, and it needs no name instrument.

6. **Router cross-validation.** The nine-route check in §5.2 is a sanity test, not a validation. A 50–100 route comparison against an independent routing source, reporting MAE, median absolute error and 90th-percentile error stratified by borough, is needed before the commute layer can be called validated.

7. **Human coding of information withholding.** §6.2 detection is keyword-based. The observed rate is 0.0% across 6,631 responses, but a blind double-coding of 100–200 sampled responses with reported agreement would establish that the rule is not simply failing to fire.

8. **Version-of-record updates.** Author names were verified on 7 September 2026 (see References note). Samad et al. (2026) is listed as appearing at AIES '26 and its proceedings pagination is not yet available; any preprint that reaches a peer-reviewed venue before final submission should be recited to the version of record.

9. **ACS covariates — completed.** The Census API key was obtained and the covariates retrieved for all 2,327 New York tracts (`code/17_fetch_acs.py`). This closed the two items previously listed here: the ACS borough column of Table 2b, and the S3 neighborhood-exposure measure, now reported in §8.11. Three notes for anyone reproducing this:

   *Failure modes are silent.* An unauthenticated request returns HTTP 200 and redirects to an HTML page titled "Missing Key"; an unactivated key returns HTTP 200 with "Invalid Key". Neither raises an exception and neither is distinguishable by status code. Check the response body.

   *The key must be activated by email link*, and there is a delay of a minute or two after activation before it is honoured. An immediate retry may fail where a retry ninety seconds later succeeds.

*The key travels in the query string*, so an unguarded traceback will print it: `urllib` includes the full URL in its exception messages. `code/17_fetch_acs.py` redacts the key from every exception it re-raises for this reason.

10. **Open-weight arm.** A fixed-weight model so at least one result remains reproducible after snapshot deprecation.

11. **Registration.** Any additional data collection should be registered before it begins (§7.6).